\documentclass[aps,prb,reprint,showpacs,showkeys,superscriptaddress,preprintnumbers,amsmath,amssymb,twocolumn,longbibliography]{revtex4-2}
\usepackage[utf8]{inputenc}
\usepackage{booktabs}
\usepackage{multirow}
\usepackage{graphicx}
\usepackage{bm}
\usepackage{mathtools}
\usepackage{dcolumn}
\usepackage{picture}
\usepackage{physics}
\usepackage{appendix}
\usepackage[colorlinks=true,urlcolor=blue,linkcolor=blue,citecolor=blue]{hyperref}
\usepackage{etoolbox}
\usepackage{natbib}
\usepackage{amsmath}
\usepackage{times}
\usepackage{color}
\usepackage{xcolor}
\usepackage{soul}
\usepackage{amssymb}
\usepackage{ulem}
\usepackage{hyperref}
\usepackage{url}
\usepackage{xr}
\usepackage[english]{babel}
\usepackage[autostyle]{csquotes}
 \DeclareUnicodeCharacter{2212}{-}
 \DeclareUnicodeCharacter{2032}{'}

\newcommand{\la}{La$_4$Ni$_3$O$_8$}
\newcommand{\nd}{Nd$_4$Ni$_3$O$_8$}
\newcommand{\pr}{Pr$_4$Ni$_3$O$_8$}

\newcommand{\T}{$\rm T_N^\ast$}
\newcommand{\etal}{\textit{et~al.}}

\begin{document}
\raggedbottom
	\title{Investigating the cause of crossover from charge/spin stripe insulator to correlated metallic phase in layered T' nickelates - R$_4$Ni$_3$O$_8$}
	
	\author{Dibyata Rout}
	\affiliation{Department of Physics, Indian Institute of Science Education and Research, Pune, Maharashtra-411008, India}
	
	\author{Sanchayeta Ranajit Mudi}
	     \affiliation{Department of Physics, Indian Institute of Science Education and Research, Pune, Maharashtra-411008, India}

       \author{Suman Karmakar}
	     \affiliation{UGC-DAE Consortium for Scientific Research, University Campus, Khandwa Road, Indore 452001}

        \author{Rajeev Rawat}
	      \affiliation{UGC-DAE Consortium for Scientific Research, University Campus, Khandwa Road, Indore 452001}
	
	\author{Surjeet Singh}
	\email[email:]{surjeet.singh@iiserpune.ac.in}
	\affiliation{Department of Physics, Indian Institute of Science Education and Research, Pune, Maharashtra-411008, India}

`   \begin{abstract}
    The T' infinite layered nickelates have recently garnered significant attention owing to the discovery of superconductivity in hole-doped RNiO$_2$ (R $=$ La, Pr, or Nd), which is the $n = \infty$ member of the series R$_{n+1}$Ni$_n$O$_{2n+2}$. Here, we investigate the $n = 3$ member, namely R$_4$Ni$_3$O$_8$ R = La, Pr, or Nd) of this family. The compound \la~exhibits simultaneous charge/spin-stripe ordering at \T~= 105~K, which is also concomitant with the onset of metal-to-insulator (MIT) transition upon lowering the temperature below \T. We investigate the conspicuous absence of this transition in the Pr and Nd analogues of \la. To achieve this purpose, we synthesized solid-solutions of the form (La, Pr)$_4$Ni$_3$O$_8$ and (La, Nd)$_4$Ni$_3$O$_8$ and examined the behavior of \T~as a function of the average R-site ionic radius ($\rm r_{\Bar{R}}$). We show that after an initial quasilinear decrease with decreasing $\rm r_{\Bar{R}}$, \T~suddenly vanishes in the narrow range $\rm 1.134~\AA~\leq~r_{\Bar{R}}~\leq~1.143~\AA$. In the same range, we observed the emergence of a new transition below $\rm T^\ast$, whose onset temperature increases as $\rm r_{\Bar{R}}$ further decreases. We, therefore, argue that the sudden vanishing of charge/spin-stripe/MIT ordering upon decreasing $\rm r_{\Bar{R}}$ is due to the appearance of a \textit{new} competing phase. The point $\rm r_{\Bar{R}} \approx r_c$, where \T~vanishes and $\rm T^\ast$ appears---a quantum critical point---should be investigated further. In this regard, \pr~and Pr-rich samples should be useful due to the weak magnetization response associated with the Pr-sublattice, as shown here.

    \end{abstract}
 
  \maketitle
  
  \section{Introduction}
	\label{Introduction}
	With the discovery of high-temperature superconductivity in cuprates by Bednorz and M\"{u}ller~\cite{Bednorz1986} in 1986, enormous efforts are underway to find superconductivity in other transition metal based oxide systems. In this regard, the most obvious place to look for is the nickel-based transition metal oxides or nickelates, isostructural and isoelectronic to the high Tc cuprates. Almost three decades ago, Anisimov \etal~\cite{Anisimov} theoretically predicted that if Ni$^{1+}$ (S $=$ 1/2) is forced into square-planar coordination, an antiferromagnetic and insulating ground state should result, which can be further hole-doped to realize superconductivity in analogy with the high Tc cuprates. The recent discovery of superconductivity in the thin films of hole doped NdNiO$_2$~\cite{Li2019}, LaNiO$_2$~\cite{Ariando_LaNiO2, Hwang_RNiO2} and PrNiO$_2$~\cite{Osada2020_PrNiO2, PrNiO2_PRM} reinvigorated the study of nickelates. The compounds RNiO$_2$ are the n $=$ $\infty$ members of a much broader infinite-layer nickelate family with the general formula of R$_{n+1}$Ni$_n$O$_{2n+2}$, where R is either an alkaline earth or a rare-earth ion and $n$ can take values 1, 2, 3,....,$\infty$). These infinite-layer nickelates can be variously located on the cuprate phase diagram depending on the Ni \textit{d} electron count, which makes them useful in realizing the different types of phases or ground states previously reported for the cuprates~\cite{Pr438_DFT_PRM, Zhang2017}. The validity of this phenomenology is only further reinforced with the discovery of superconductivity in thin films of \enquote{quintuple - layer} compound Nd$_6$Ni$_5$O$_{12}$ (n $=$ 5)~\cite{Pan2022}, which falls in the optimally doped region of the cuprate phase diagram~\cite{Keimer2015}.

\begin{figure*}[t]
	\centering
	\includegraphics[trim=3.8cm 0.7cm 3.8cm 0.4cm,clip=true,width=15cm]{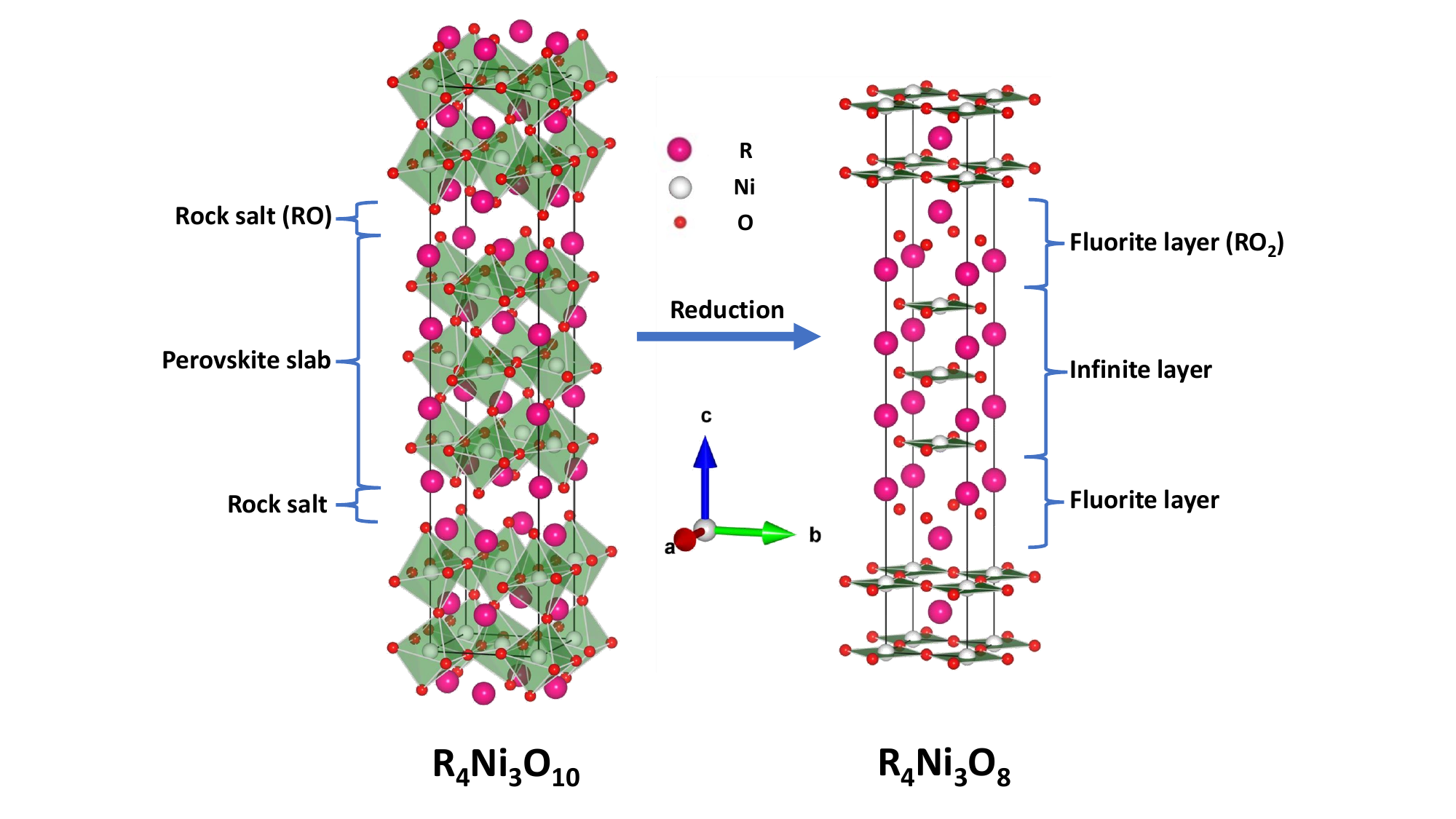}
	\caption {The crystal structure of the n = 3 member of the parent Ruddlesden-Popper nickelate phase (left) and their reduced T' variants (right).}
	\label{CS}
    \end{figure*}

    In this work, we focus our attention on the n $=$ 3 member of the infinite layer nickelate family i.e., R$_4$Ni$_3$O$_8$ (where R $=$ La, Pr and Nd). They crystallize in the tetragonal space group $I4$$\slash$$mmm$ (No. 139) with three infinite NiO$_2$ planar layers  separated by an intervening fluorite layer (RO$_2$) as shown in Fig.~\ref{CS} (right panel). The R$_4$Ni$_3$O$_8$ compounds are mixed-valent, containing Ni$^{1+}$/Ni$^{2+}$ in the ratio of 2:1 which resemble the 3d$^9$/3d$^8$ electronic configuration of Cu$^{2+}$/Cu$^{3+}$ present in high Tc cuprates. The average Ni valence in R$_4$Ni$_3$O$_8$ is +1.33 i.e., a \textit{d} filling value of 8.67, which essentially lies in the overdoped, Fermi liquid regime of the cuprate phase diagram~\cite{Zhang2017natphy}. This, 
    and their similarities with cuprates, including a large orbital polarization of the unoccupied \textit{e$_g$} states~\cite{Zhang2017}, strong Ni 3$d$ and O 2$p$ hybridization~\cite{superexchange_PRL2021}, and the square-planar arrangement of Ni are reasons enough to investigate their physical properties at low temperatures. 

    \la ,~\pr ~and \nd~exhibit contrasting ground state properties despite belonging to the same crystallographic space group. While \la~ is a charge/spin (CS) stripe-ordered insulator which features a sharp semiconductor-to-insulator transition (henceforth, we shall loosely refer to it as a metal-to-insulator or MIT transition) concomitant with onset of charge/spin-stripe ordering below at temperature 105~K, which we shall designate as $\rm T_N^\ast$ in the rest of the manuscript~\cite{PNASZhang, PhysRevLett.104.206403, PRLPressureLa, PRL_NPD_La, PhysRevB.83.014402_NMR}, Pr$_4$Ni$_3$O$_8$ and \nd~are reported to show a metallic behavior over the whole temperature range with no indications of charge-stripe or spin ordering~\cite{Zhang2017natphy, miyatake4080321chemical, Nakata2016, Pan2022}
    Understanding these stark differences in the ground state properties as a function of R-site ionic radius, and a careful analysis of the magnetic properties of the R-sublattice in Pr and Nd analogues, which has not been not well explored in the past, are important. However, these compounds require a part of Ni to be in an unfavourable oxidation state of $+1$, which makes their formation in pure phase somewhat challenging, hence limiting a detailed investigation of their physical properties.
    
    In this study, we investigate the ground state of \la,~\pr~and~\nd ~. At the same time, to understand the role of R-site ionic radius on the concomitant CS stripe ordering in \la, we investigate solid-solutions of the form (La$_{1-x}$R$_x$)$_4$Ni$_3$O$_8$ (R $=$ Pr, Nd) in well-characterized samples. 
    A phase diagram showing the evolution of~\T~with average R-site ionic radius, $\rm r_{\Bar{R}}$, has been constructed by combining data from the (La$_{1-x}$Pr$_x$)$_4$Ni$_3$O$_8$ and (La$_{1-x}$Nd$_x$)$_4$Ni$_3$O$_8$ series for various $x$. We show that \T~decreases nearly linearly upon decreasing $\rm r_{\Bar{R}}$ at first, and this trend continues down to about $\rm r_{\Bar{R}} = 1.143~\AA$ after which \T~drops sharply to zero. We show that at the critical average R-site ionic radius ($\rm r_c$) where \T~goes to zero, a new transition emerges, denoted as T$^\ast$, which shows an almost linearly increasing trend with decreasing $\rm r_{\Bar{R}}$ in the region $\rm r_{\Bar{R}} < r_c$. We, therefore, associate the sudden disappearance of charge/spin-stripe order upon dilution at the La-site with Pr or Nd, with the emergence of a new competing phase as $\rm r_{\Bar{R}}$ decreases below a certain critical $r_c$.
    The rest of the paper is organized as follows: The details of the experimental methods are given in Sec.~\ref{Expt}, followed by results and discussion in Sec.\ref{Results and Discussion}. The details of sample synthesis, crystal structure and low-temperature synchrotron x-ray diffraction appear in Sec. \ref{STR}. The electrical transport, magnetic susceptibility and specific heat data are discussed in Sec. \ref{PHYC}. The summary and conclusions drawn are presented under Sec. \ref{SC}.

    \begin{figure*}
		\centering
		\includegraphics[width= 2\columnwidth]{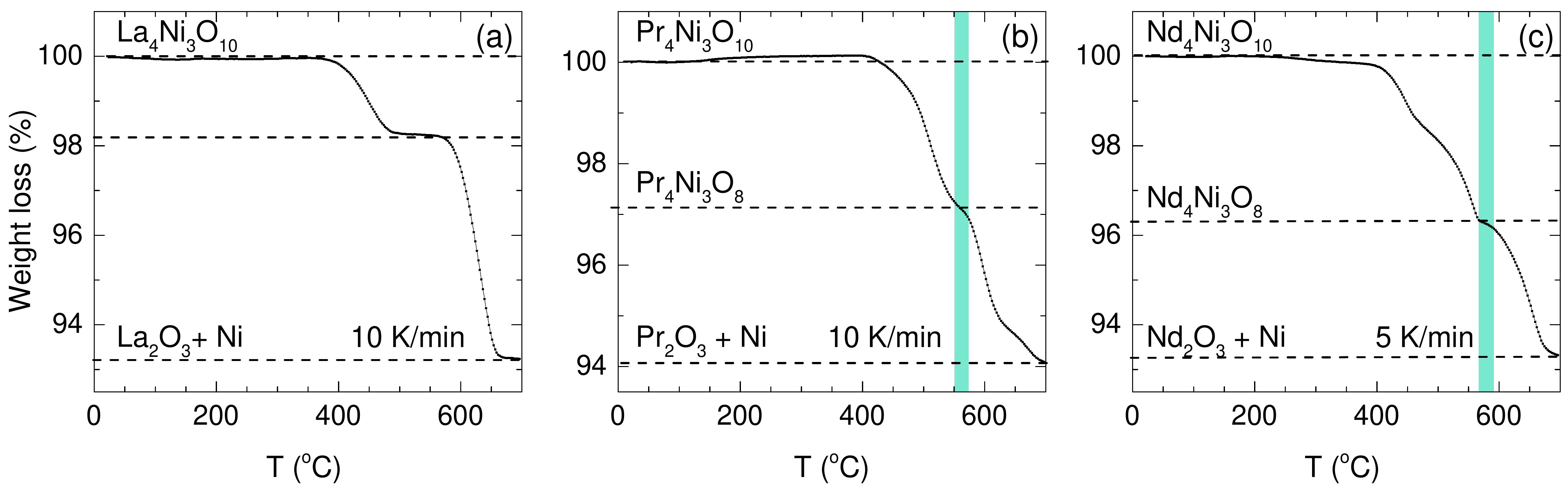}
		\caption {(a), (b) and (c) show the TGA of La$_4$Ni$_3$O$_{10}$, Pr$_4$Ni$_3$O$_{10}$ and Nd$_4$Ni$_3$O$_{10}$ respectively, carried out in Ar - H$_2$ (10\%) atmosphere to determine the synthesis protocol for R$_4$Ni$_3$O$_8$ samples. The aqua-blue rectangle marks the plateau region where R$_4$Ni$_3$O$_8$ phase (R $=$ Pr and Nd) is stabilized.}
		\label{TGALL}
    \end{figure*}

    \section{Experimental Methods}
	\label{Expt}
	High purity polycrystalline samples of the n $=$ 3 member of the parent Ruddlesden Popper (RP) phases i.e., (La$_{1-x}$R$_x$)$_4$Ni$_3$O$_{10}$ (R $=$ Pr, Nd; x $=$ 0, 0.1, 0.5, 0.75, 0.9 and 1.0) were prepared using the citrate method as described in~\cite{Dibyata2020}. 
    The phase purity of these samples was confirmed using a Bruker D8 Advance powder X-ray diffractometer (PXRD). After this, the corresponding trilayer T' nickelates were obtained by either the topotactic reduction using CaH$_2$ or reduction under a stream of H$_2$ gas (for \la) or Ar/H$_2$ mixture (for Pr, Nd mixed samples). The chemical composition of the samples was analyzed using the energy-dispersive X-ray analysis (EDX) technique in a Zeiss Ultra Plus scanning electron microscope. To confirm the oxygen stoichiometry of our samples, we carried out the complete decomposition of the formed samples under $10\%$ Ar-H$_2$ atmosphere, employing a heating rate of 10 K min$^{-1}$ in a high-resolution TGA setup (Netzsch STA $449$ F1).

    \begin{figure*}
		\centering
		\includegraphics[trim=0.0cm 0cm 0.6cm 0cm,clip=true,width= 1.5\columnwidth]{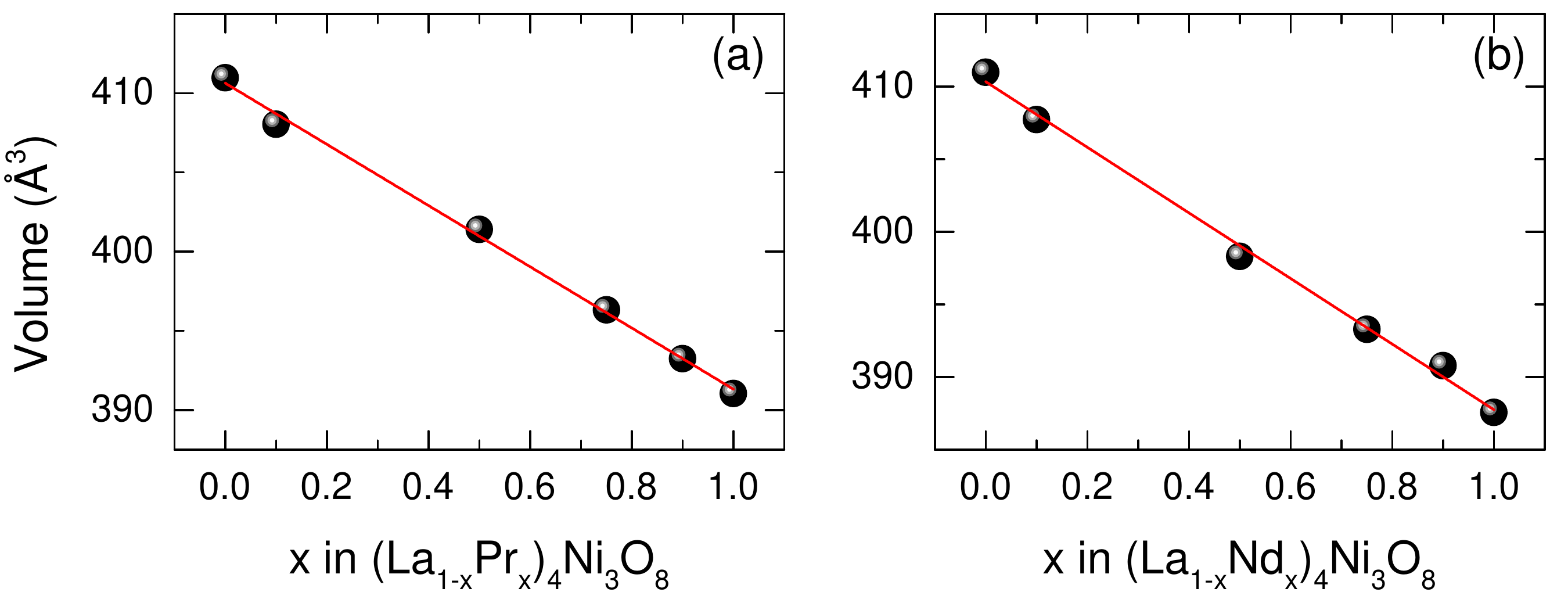}
		\caption {(a) and (b) show the variation of unit cell volume for La$_{1-x}$Pr$_x$Ni$_3$O$_8$ and La$_{1-x}$Nd$_x$Ni$_3$O$_8$ (x $=$ 0, 0.1, 0.5, 0.75, 0.9 and 1.0) samples respectively, estimated using lab-based PXRD. The solid red line denotes a linear fit to the data.}
		\label{XRD}
     \end{figure*}

     \begin{figure*}
		\centering
		\includegraphics[width= 1.9\columnwidth]{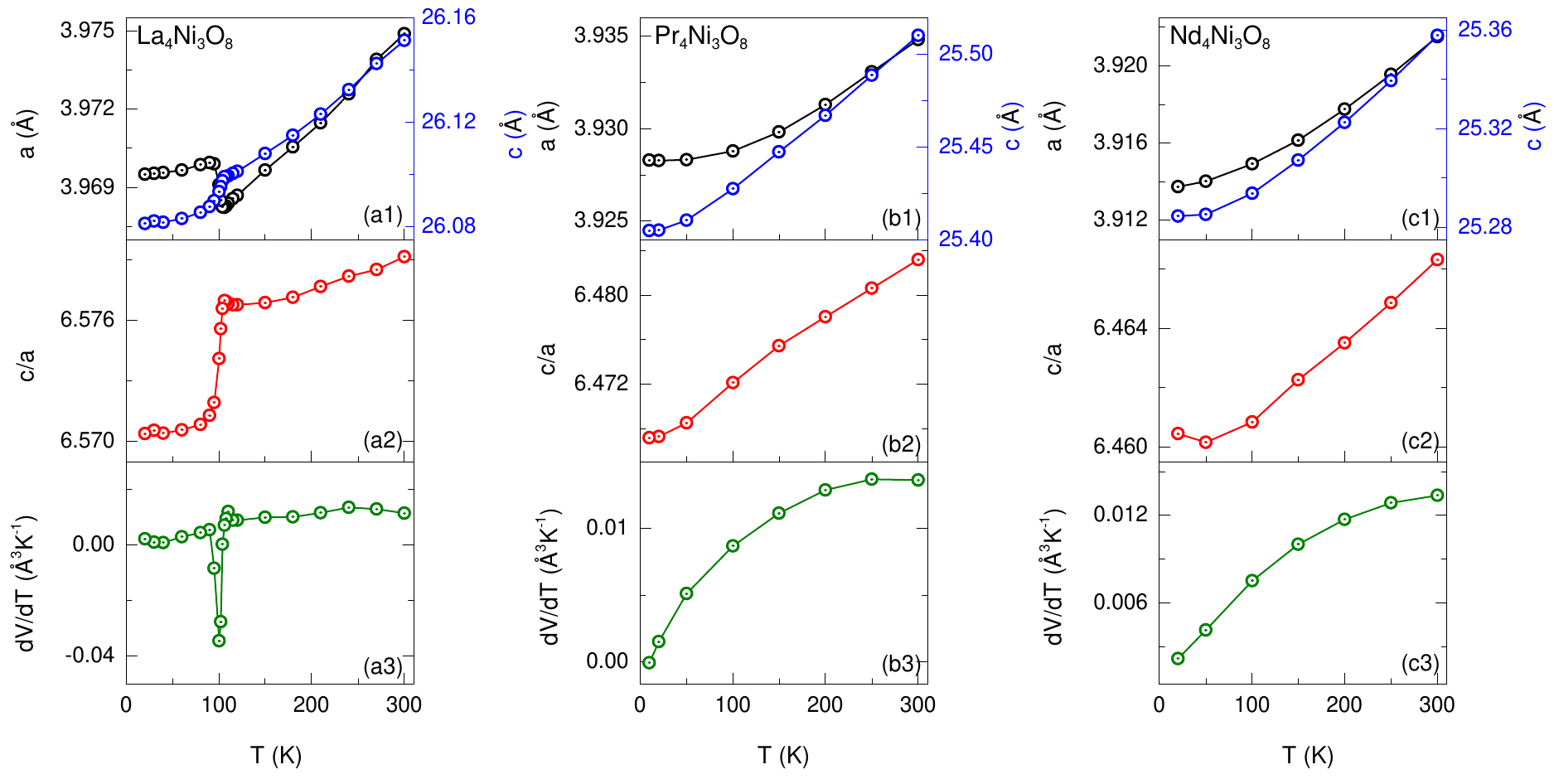}
		\caption {(a1-a3), (b1-b3) and (c1-c3) show the temperature variation of lattice parameters, c/a ratio and dV/dT for the \la~, \pr~ and \nd~ sample respectively. 
  }
		\label{XRD1}
    \end{figure*}

    High-resolution synchrotron powder X-ray diffraction experiments were carried out at the MSPD-BLO4 beamline of the ALBA synchrotron center, Barcelona, Spain. The samples were prepared in the form of finely ground powders that were placed in a borosilicate capillary tube of $0.5$ mm inner diameter. For cooling the sample down to 90 K, an Oxford Cryostream $700$ series nitrogen blower was used while for attaining temperatures lower than 90 K, He exchange gas was used. The diffractograms were collected in the range $0^o\leq 2\theta\leq30^o$ with a step size of $0.003^o$. The incident beam energy was set at $38$ keV ($\lambda =0.3263$~\AA) and a high-resolution detector (MAD$26$) was used to resolve any subtle structural modifications. On the other hand, the low temperature lab-based PXRD measurements were performed on Malvern Panalytical Empyrean Series 3 X-ray diffractometer with Oxford Cryosystems PheniX closed-cycle helium cryostat, which can attain a lowest possible temperature of 12 K. The structural refinement was done by the Rietveld method using the FULLPROF suite \cite{rodriguez1993recent}.

    Heat capacity, magnetization and resistivity measurements were carried out using the Physical Property Measurement System (PPMS), Quantum Design USA
    . The heat capacity of the sample holder and APIEZON N grease (addenda) was determined prior to the measurements. The magnetization measurements were carried out in the zero-field cooled (ZFC) mode at an applied field of $H = 90$~kOe. Isothermal magnetization was carried out at a temperature of 5 K up to an applied field of 90~kOe. Resistivity measurements were carried out on rectangular bar samples of known dimensions using the standard four-probe technique.
    
    High-resolution transmission electron microscopy (HRTEM) was carried out using a JEOL JEM 2200FS 200keV TEM instrument. The powder samples were finely ground in high-purity ethanol using an agate mortar and pestle to reduce the formation of agglomerates. Thereafter, less than a few mg of the ground powder was dispersed in an ethanol solution and was subjected to sonication for a period of 30 min. A few droplets of the resultant suspension were drop-casted onto a TEM Cu-grid using a micropipette. The Cu grid was then dried for 12 h in an evacuated desiccator, preheated at 60$^\circ$C in an oven for 15 min, and eventually loaded into the TEM sample chamber. Both HRTEM and SAED (Selected Area Electron Diffraction) patterns were collected for all the samples and the analysis of the images was carried out using DigitalMicrograph (GMS -3) software package.

    \begin{figure*}
		\centering
		\includegraphics[width=14cm]{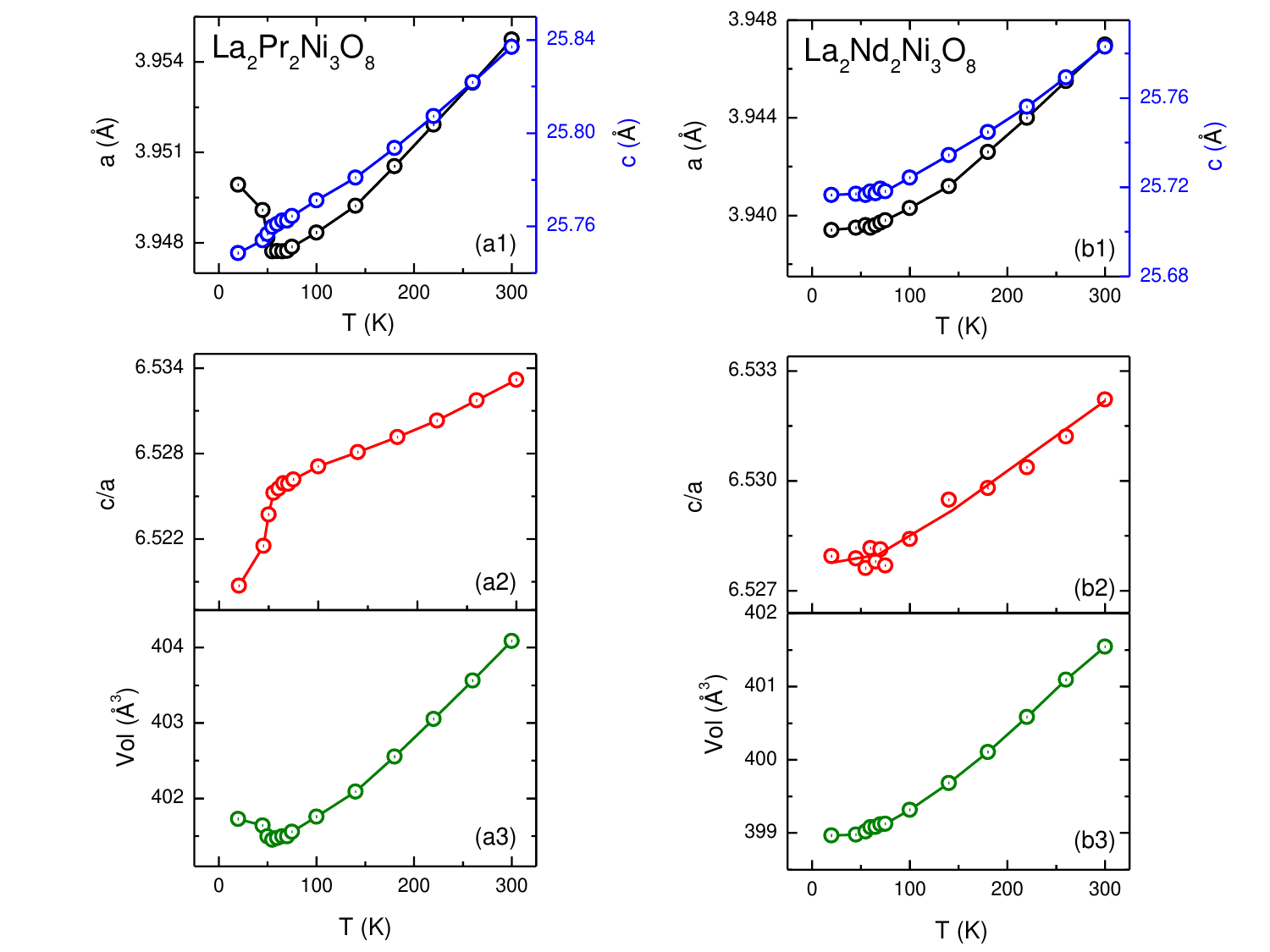}
		\caption {(a1 - a3) and (b1 - b3) show the temperature variation of lattice parameters, c/a ratio and unit cell volume for the La$_2$Pr$_2$Ni$_3$O$_8$ and La$_2$Nd$_2$Ni$_3$O$_8$ sample respectively, measured from room temperature down to the lowest attainable temperature of 10 K.}
		\label{XRD2}
    \end{figure*}

    \begin{figure*}
		\centering
		\includegraphics[width= 1.4\columnwidth]{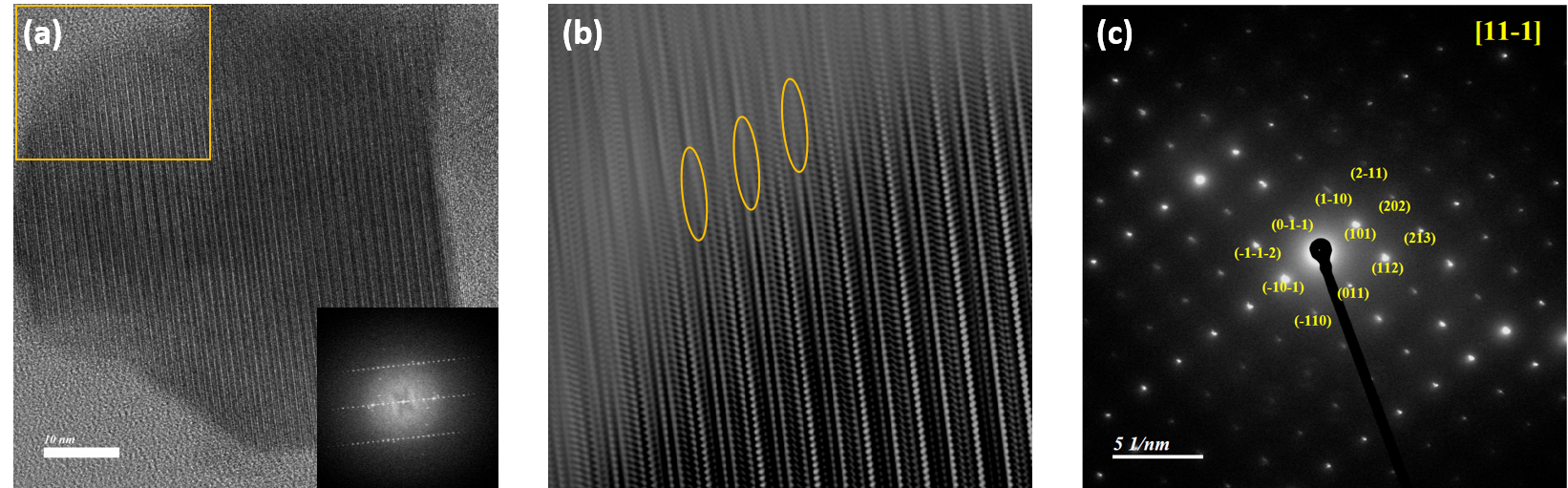}
		\caption {(a) HRTEM micrograph of Pr$_4$Ni$_3$O$_8$; inset at right bottom shows the FFT image of the micrograph. (b) IFFT of the region inside the yellow box shown in (a), showing the presence of stacking faults in the sample; (c) shows the SAED pattern taken on a highly crystalline region of the specimen, along the [1 1 -1] zone axis with the $hkl$ indices marked in yellow. }
		\label{TEM}
    \end{figure*}

     \section{Results and Discussion}
     \label{Results and Discussion}
     \subsection{Sample synthesis}
     \label{STR}
     As is mentioned in the experimental section, the R$_4$Ni$_3$O$_{8}$ samples were prepared by reducing the parent RP phases. To obtain optimal conditions for the reduction, the parent RP phases were first decomposed completely in a TGA setup under Ar-H$_2$ (10\%) atmosphere as shown in Fig.~\ref{TGALL}. From these TGA plots, we conclude that (Pr/Nd)$_4$Ni$_3$O$_8$ samples and the (La$_{1-x}$R$_x$)$_4$Ni$_3$O$_8$ samples for $x > 0.5$, can be obtained by heating at the sample at 360$^\circ$C for a duration ranging from 19~h to 22~h. Moreover, this reaction time could be further reduced by increasing the isotherm temperature to 500$^\circ$C for 1.3~h for \pr~and 470$^\circ$C for 5~h for \nd, respectively. However, for the reductions at these higher temperatures, the gas flow had to be changed from Ar - H$_2$ (10\%) to pure Ar, immediately at the end of the isotherm via a T-connector. 
     
     On the other hand, as is evident from the TGA data in Fig.~\ref{TGALL}(a), \la~could not be obtained under Ar-H$_2$ (10\%) atmosphere as the step expected near 3\% weight loss is missing from the plot. On the contrary, the step is present near 2\% weight loss, which signals the appearance of a new phase. This is also in accordance with the TGA data by Laccore \textit{et al}.~\cite{LACORRE1992495}. The synthesis of \la~was therefore optimized under ultra-high purity grade (UHP) H$_2$ gas. 
     An isotherm temperature of 470$^\circ$C for 28~min and sample mass ranging from 40-50 mg are found to be ideal conditions for obtaining pure \la~phase. A slight increment in the isotherm duration resulted in the decomposition of the sample to La$_2$O$_3$ and Ni metal. Interestingly, we also captured a new phase for the same sample mass and isotherm temperature but with the isotherm duration in the range of 20-25 min (see Fig. S2 in the Supplementary Materials). This new phase either corresponds to the La$_4$Ni$_3$O$_9$ phase previously alluded to by Laccore~\textit{et al}.~\cite{LACORRE1992495}, or it could be related to the T$^\dagger$ phase of \la~reported by Cheng~\textit{et al}.~\cite{PRLPressureLa}. Further detailed structural characterization experiments are needed to ascertain the exact crystal symmetry of this new phase. Alternatively, we also synthesized the \la~compound, and 10\% (Pr/Nd) doped \la~samples, using the method of topotactic reduction employing CaH$_2$ as a reducing agent. In this case, a sintered rectangular bar of the parent RP phase was covered in an optimized amount of CaH$_2$ powder in a quartz ampoule in an argon-filled glove box. The quartz ampoule was then flame-sealed under vacuum (10$^{-5}$ Torr) and subjected to an isotherm temperature of 360$^\circ$C for a period of 36~h.\\ 
     
     \subsection{Structural Characterization}
     The room temperature crystal structure of R$_4$Ni$_3$O$_{8}$ (R $=$ La, Pr and Nd) is shown in Fig.~\ref{CS} (right panel). The variation of unit cell volume 
    with Pr or Nd doping $x$ for the (La$_{1-x}$Pr$_x$)$_4$Ni$_3$O$_8$ and (La$_{1-x}$Nd$_x$)$_4$Ni$_3$O$_8$ samples is shown in Fig.~\ref{XRD}(a) and Fig.~\ref{XRD}(b), respectively. In both cases, a monotonic decrement of the unit cell volume with increasing $x$ is observed, confirming the successful doping at the La-site. 
     During the reduction from R$_4$Ni$_3$O$_{10}$ to R$_4$Ni$_3$O$_8$, the apical oxygen atoms are removed from the NiO$_6$ octahedra present in the perovskite slab in R$_4$Ni$_3$O$_8$. Thus, the octahedral arrangement around the Ni atoms in the parent RP phase changes to a square-planar arrangement in the T' phase, transforming the perovskite trilayer block into an infinite-trilayer block comprising a stack of three infinite or planar NiO$_2$ layers. Henceforth, we shall refer to this block as a planar-trilayer or infinite-trilayer or quite simply as a trilayer-block when there is no ambiguity. Similarly, the original rock salt layer in the RP structure transforms into a fluorite-type RO$_2$ layer upon reduction. This fluorite RO$_2$ layer acts as a buffer between any two successive trilayer blocks. The T' structure has two distinct rare-earth sites denoted by R1 and R2, where R1 lies in the trilayer block and R2 faces the fluorite layer on one side and the trilayer block on the other. Likewise, there are two distinct crystallographic sites for the Ni atoms denoted by Ni1 and Ni2, where Ni1 lies within the trilayer block, and Ni2 is sandwiched between the fluorite and trilayer blocks.
     
     Fig.~\ref{XRD1} shows the temperature variation of lattice parameters of the parent R$_4$Ni$_3$O$_{8}$ (R $=$ La, Pr and Nd) samples down to 10~K. The temperature variation of the lattice parameters $a$ and $c$ of \la~is shown in Fig.~\ref{XRD1}(a1). The corresponding plots showing the temperature variations of $c/a$ and $\rm dV/dT$, where $\rm V$ is the unit cell volume, are shown Fig.~\ref{XRD1}(a2) and Fig.~\ref{XRD1}(a3), respectively. The lattice parameters exhibit a clear anomaly at T $\sim$ 105 K, which coincides with the temperature \T~where the MIT or CS-stripe order is expected to set in on the basis of previous studies. Below \T, the parameter $a$ shows an anomalous increase with decreasing temperature; the parameter $c$ on the other hand shows a step-like decrease; and accordingly, the ratio $c/a$ presents a sharp, step-like decrease at \T. These variations are in agreement with previous reports \cite{PRLPressureLa, Zhang2017}. 
     Contrary to \la, the lattice parameters in \pr, Figs.~\ref{XRD1}(b), and \nd, Fig.~\ref{XRD1}(c), show a smooth and monotonic decrease upon cooling down to the lowest measured temperature of 10~K. 
     
     Coming now the temperature variation of lattice parameters of the doped samples. The lattice parameters of La$_2$Pr$_2$Ni$_3$O$_8$ and La$_2$Nd$_2$Ni$_3$O$_8$ are shown in Fig.~\ref{XRD2}. In La$_2$Pr$_2$Ni$_3$O$_8$, the 105~K anomaly suppresses down to a temperature of 55~K. The qualitative behaviour of $a$ and $c$ across this transition remains similar to that described above for~\la, but the anomaly at \T~has weakened considerably, as shown in Fig.~\ref{XRD2}(a). However, in La$_2$Nd$_2$Ni$_3$O$_8$, \T~suppresses completely, with the lattice parameters showing a monotonically decreasing behaviour down to the lowest temperature in our measurement, as shown in the right panels of Fig.~\ref{XRD2} where the temperature variation of $a$, $c$, $c/a$ and V are shown. This difference (i.e., La$_2$Pr$_2$Ni$_3$O$_8$ showing a transition but La$_2$Nd$_2$Ni$_3$O$_8$ not) can be attributed to the ionic radii difference ($\rm r_{Pr} > r_{Nd}$). The smaller size of Nd induces a larger negative chemical pressure compared to the same amount of Pr, and because the decreasing average R-site radius ($r_{\Bar{R}}$) reduces \T, it is expected that below a certain critical $r_{\Bar{R}} = r_c$ the transition will disappear. In other words, as $\rm r_{\Bar{r}} \rightarrow r_c$ from above ($\rm r_{\Bar{r}} > r_c$),  \T~$\rightarrow 0$. The value of $\rm r_c$ has been recently shown to lie in the range $\rm 1.136~\AA \leq r_{\Bar{R}} \leq 1.140~\AA$~\cite{Xinglong_CM}. The average ionic radius of R = La$_{0.5}$Nd$_{0.5}$ ($\rm r_{\Bar{R}}$ = 1.1345~\AA) is clearly on the lower side of this range; whereas for R = La$_{0.5}$Pr$_{0.5}$ ($\rm r_{\Bar{R}}$ = 1.143~\AA), the average ionic radius is clearly on the higher side of this range and hence \T~is non-zero for this sample. 
     The temperature-dependent XRD data for the 10\% Pr and Nd samples, collected using a lab-based diffractometer are shown in Fig. S3 of the Supplementary Material. For both doping types, \T~suppresses to 99~K and 92~K for Pr-and Nd-doped samples, respectively. 
              
     Fig.~\ref{TEM}(a) shows the HRTEM micrograph of \pr~sample. The sample shows nicely lined up crystal planes with few defects in the form of stacking faults as shown in yellow ellipses in Fig.~\ref{TEM}(b) where the Fast Fourier transform (FFT) image of the region enlosed within the yellow-colored border in Fig.~\ref{TEM}(a) is shown. No sign of intergrowth due to lower and higher $n$ members, which typically plagues the sample quality, could be seen in our samples. The SAED pattern, shown in Fig~\ref{TEM}(c), consists of sharp spots indicative of a high crystallinity of the sample. The SAED pattern satisfies the reflection conditions expected for the $I4/mmm$ space group in agreement with the previous HRTEM studies on \nd~\cite{RETOUX1998307}.
     
    \subsection{Physical characterization}
    \label{PHYC}
    \subsubsection{Magnetization}
    
    \begin{figure*}[tbh!]
		\centering
		\includegraphics[width=1.8\columnwidth]{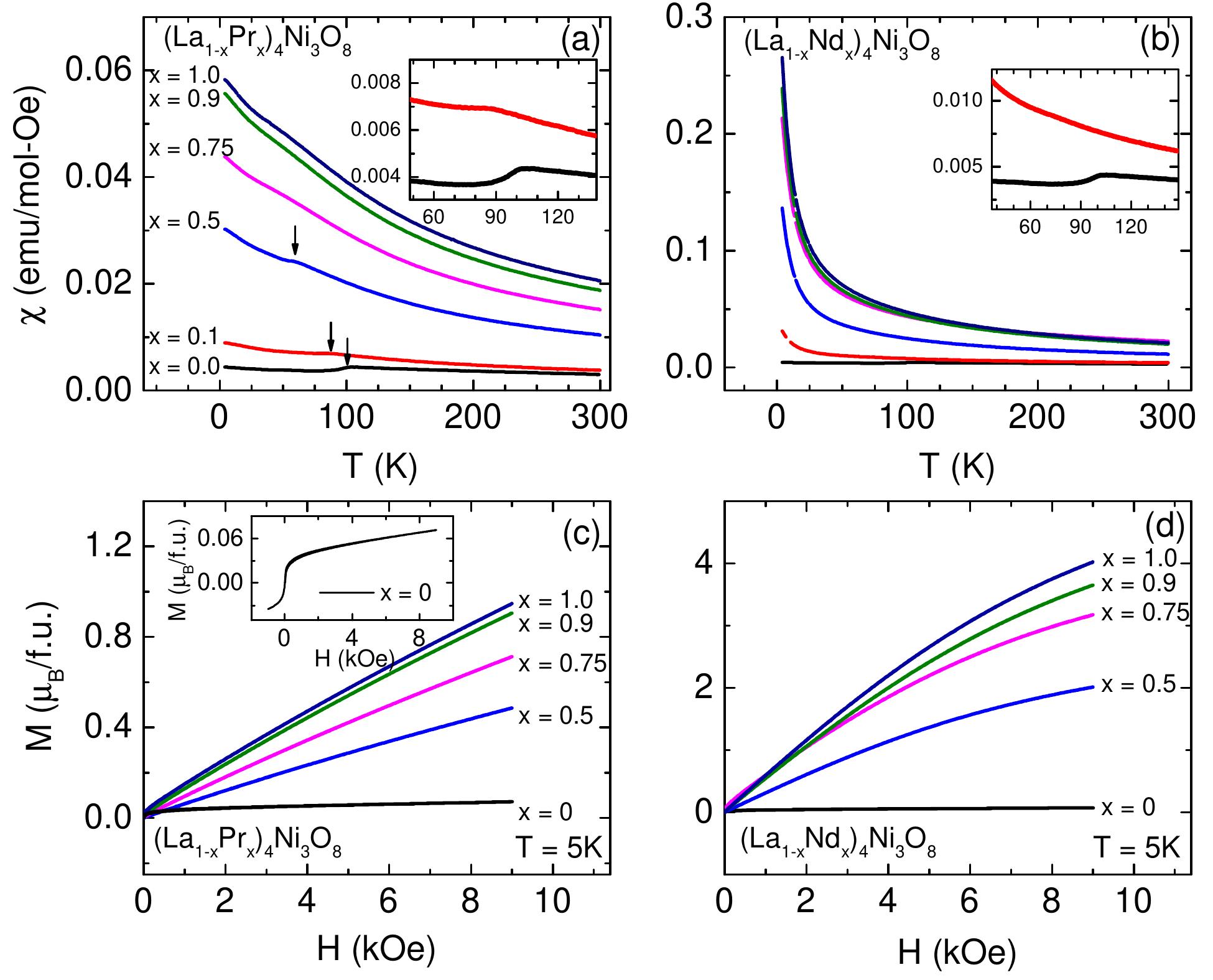}
		\caption {(a) and (b) show the zero field cooled magnetic susceptibility as a function of temperature for the La$_{1-x}$Pr$_x$Ni$_3$O$_8$ and La$_{1-x}$Nd$_x$Ni$_3$O$_8$ (x $=$ 0, 0.1, 0.5, 0.75, 0.9 and 1.0) samples respectively at an applied field of 90 kOe; (c) and (d) show the isothermal magnetization at T $=$ 5 K for the Pr and Nd series respectively measured from 0 to 90 kOe.}
		\label{MT}
    \end{figure*}

    Fig.~\ref{MT}(a) and Fig.~\ref{MT}(b) show the normalized magnetization, M/H, as a function of temperature for the (La$_{1-x}$Pr$_x$)$_4$Ni$_3$O$_8$ and (La$_{1-x}$Nd$_x$)$_4$Ni$_3$O$_8$ samples, respectively, where H is the applied field in kOe units. In \la~ (x = 0), the concomitant charge/spin-stripe ordering is seen as a kink in the M/H plot near 105~K. A zoomed-in view of this feature is shown in the inset in Fig.~\ref{MT}(a), where it can be seen more clearly. The temperature at which this anomaly appears in M/H is in good agreement with the previous reports~\cite{PNASZhang, PhysRevLett.104.206403, PRLPressureLa}. 
    Upon doping with Pr or Nd the transition is suppressed. However, an important difference between the two series of samples is that the magnetization of the Nd series at low temperatures is significantly high compared to the corresponding Pr series for any given $x$. For example, for \pr~, M/H at T $=$ 5 K is about $\rm \approx 6 \times 10^{-2}~emu~mol^{-1}~Oe^{-1}$ whereas in \nd, M/H at the same temperature is $\rm \approx 26 \times 10^{-2}~emu~mol^{-1}~Oe^{-1}$; therefore, the magnetization of~\pr~is almost 4 to 5 times reduced compared to that of~\nd~. This difference is also reflected in the isothermal magnetization plots shown in Fig.~\ref{MT}(c) and Fig.~\ref{MT}(d). Since the calculated effective magnetic moment of a free Pr$^{3+}$ ion is not very different from that of Nd$^{3+}$ (3.58~$\rm \mu_B$ and 3.62~$\rm \mu_B$, respectively), the reduced values of M/H of the Pr sample at low temperatures is likely a manifestation of the crystal field splitting of the lowest $J$-multiplet ($J = 4$) of the Pr$^{3+}$ ions. Previously, we observed a similarly reduced M/H at low temperatures for Pr$_4$Ni$_3$O$_{10}$ as compared to Nd$_4$Ni$_3$O$_{10}$ due to the non-magnetic-singlet ground state associated with one of the two crystallographically inequivalent Pr ions in Pr$_4$Ni$_3$O$_{10}$~\cite{Dibyata2020}. We believe that a similar scenario prevails in the T' analogue~\pr. As a result, in (La$_{1-x}$Pr$_x$)$_4$Ni$_3$O$_8$ series, the low-temperature magnetization signal originating from the Ni sublattice is not as heavily masked by the paramagnetic background due to the rare-earth moments as in the Nd case. Hence, in the Pr-doped series, the anomaly due to CS-stripe ordering remains clearly discernible up to x = 0.5. The transition temperature from the M/H plots (indicated by the position of arrows) is in fairly good accord with the temperature-dependent PXRD data discussed in the previous section, and with the transition temperature reported for the (La$_{1-x}$Pr$_x$)$_4$Ni$_3$O$_8$ series in Ref.~\cite{Xinglong_CM}. Beyond x = 0.5 (i.e., x = 0.75, 0.9, and 1), we continue to see the presence of a weak hump around 35~K. In Fig. S5 in the Supplementary Material, the derivative plots clearly capture the presence of this feature. It should be noted that this feature appears even in the \pr~sample (in fact, it is relatively less dominant in x = 0.75 and 0.9), which suggests that this feature is not related to the CS-stripe ordering. Since the size of this feature scales with Pr concentration, it is fair to conclude that it originates from the Pr-sublattice and is likely a manifestation of the crystal field splitting of the lowest $J$-multiplet of the Pr$^{3+}$ ions. 
    The M/H of our samples agrees fairly well with the M/H data for H $\parallel$ \textit{ab} plane in Ref.~~\cite{Xinglong_CM}. When measured with the applied field aligned parallel to the c-axis, M/H exhibits an additional hump around 100 K.

    We now investigate the magnetic behaviour of \nd. Being a Kramers' ion with $f^3$ configuration (odd number of $f$ electrons), the crystal field split ground state of Nd$^{3+}$ ($J = 9/2$) cannot be a singlet (i.e., the ground states of both Nd1 and Nd2 will be either a doublet or a quartet as the Kramers' theorem forbids a singlet ground state). Hence, at low temperatures, both Nd sites contribute to the magnetization, and the overwhelming paramagnetic background that masks the CS-stripe anomaly associated with the Ni sublattice. This is true even in the sample with as small as 10\% Nd doping (x $=$ 0.1). For higher doping values, the Curie-like behaviour dominates over the whole temperature range. 
    
    For the \pr~ and \nd~ samples, $\chi^{-1}$ vs. T plots look fairly linear with small curvatures above about 150~K.
    We, therefore, fitted the high-temperature data using the modified Curie-Weiss (CW) law: $\rm \chi = \chi_0 + C/(T - \theta_p)$, where C is the Curie-constant from which the value of the effective magnetic moment ($\rm \mu_{eff}$) can be obtained using $\rm \mu_{eff} = \sqrt{8C}$, and $\theta_p$ is the Weiss temperature, and $\chi_0$ is the temperature independent contribution arising from the core-diamagnetism and the paramagnetism of Van Vleck or Pauli type. Treating $\rm \chi_0$, $\rm C$ and $\rm \theta_p$ as the fitting parameters, the best fit over the temperature range 150~K to 300~K using the modified Curie-Weiss equation led to the following values of the fitting parameters: 
    For \pr, $\rm \chi_0 = 2.0 \times 10^{-3}~emu~mol^{-1}~Oe^{-1}$, $\rm C = 7.3~emu~mol^{-1}~Oe^{-1}~K$, and $\rm \theta_p = -96~K$. The corresponding values for \nd~are: $\rm \chi_0 = 1.6 \times 10^{-3}~emu~mol^{-1}~Oe^{-1}$, $\rm C = 6.8~emu~mol^{-1}~Oe^{-1}~K$ and $\rm \theta_p = -49~K$. If we ignore the contribution of the Ni sublattice for the time being then we should get $\rm \mu_{eff}/Pr^{3+} = 3.83 \mu_B$ in \pr, which exceeds the Hund's rule derived \textit{free-ion} value of $\rm 3.58~\mu_B$ substantially. 
    This suggests that the magnetization associated with the Ni sublattice is non-negligible. 
    Now, we know that the ratio Ni$^{1+}$/Ni$^{2+}$ in T' structure is 2:1, and Ni$^{1+}$ has a spin 1/2 while Ni$^{2+}$ can take two possible spin states: spin 0 (low-spin or LS) or spin 1 (high-spin or HS). Accordingly, the value that quantity $\rm 3(\mu_{Ni})^2$ can be $\rm 14~\mu_B^2$ (HS) or $\rm 6~\mu_B^2$ (LS) (assuming $g$ factor to be 2 in each case). We calculated value of $\rm 3(\mu_{Ni})^2$ using the relation: 
    $\rm 4(\mu_{Pr})^2 + 3(\mu_{Ni})^2 = 8C$, where $\rm \mu_{Pr} = 3.58~\mu_B$ in the free-ion moment on Pr$^{3+}$ and $\rm C = 7.3~emu~mol^{-1}~Oe^{-1}~K$. This gives a value of $\rm \approx 7 ~\mu_B^2$ for $3\mu_{Ni}^2$, which is closer to the value expected in the low-spin case, indicating that the Ni$^{2+}$ in \pr~ is in a low-spin state. In the case of \nd, using $\rm C = 6.8~emu~mol^{-1}~Oe^{-1}~K$ and $\rm \mu_{Nd} = 3.62~\mu_B$, we get $\rm 3\mu_{Ni}^2 \approx 2~\mu_B^2$, which is smaller than the value $6~\mu_B^2$ expected in the LS case, suggesting that the $d$-electrons of Ni are partly itinerant. 
    This is consistent with the experimental observation that \nd~sample is more electrically conducting compared to \pr~(\textit{vide infra}). This, rather crude, analysis of M/H above 160~K appears to indicate that Ni $d$-electrons in R$_4$Ni$_3$O$_8$ become more itinerant as the ionic radius of R$^{3+}$ decreases. The high value of $\rm \theta_p$ and $\rm \chi_0$ in both cases are likely a consequence of large crystal field splitting. It should be recalled that for an accurate estimate of $\rm \theta_p$, $\rm \chi_0$, and $\mu_{eff}$, the fittings should ideally be performed over a broad temperature range satisfying $\rm k_BT >> \Delta_{CF}$, where $\rm \Delta_{CF}$ is the overall crystal field splitting of the lowest J-multiplet. 
    
    We thence tried fitting the low-temperature data for \pr~and~\nd~to extract the magnetic moment in the crystal field split ground state level. In \nd, a satisfactory fit could be obtained between 10~K and 25~K, yielding fitting parameters as follows: $\rm \chi_0 = 0.031 \times 10^{-3}~emu~mol^{-1}~Oe^{-1}$, $\rm C = 1.96~emu~mol^{-1}~Oe^{-1}~K$ and $\rm \theta_p = -4.7~K$. We see that the Curie constant has reduced drastically from its high-temperature value suggesting that the magnetic moment per Nd ion in the crystal field split ground state is considerably smaller than the free-ion value. The values of $\chi_0$ and $\theta_p$ are now less affected by the crystal field splitting than before; however, unless the ground state is well-insulated from the excited states (which indeed in the case as we shall see), the true values, interpretable as the strength of exchange coupling ($\theta_p$) and core-diamagnetism/Pauli paramagnetism ($\chi_0$), cannot be correctly ascertained in most cases. Interestingly, in the case of~\pr, a satisfactory Curie-Weiss fit could not be obtained despite adjusting the upper and lower temperature limits of the fitting range
    , indicating the complex nature of the magnetic ground state in ~\pr. This we shall revisit as we advance further. 
    
    The isothermal magnetization at T = 5 K in both sets of doped samples scales with the mole fraction of the rare-earth element present (Fig.~\ref{MT}(c) and Fig.~\ref{MT}(d)). The temperature variation of M(H) is quasi-linear for \pr~and Pr-doped samples, but a substantial non-linearity can be seen for the \nd~and Nd-doped samples. In fact, in the Nd-doped samples, the magnetization near the highest field is on the verge of plateauing. At T = 5 K, and under an applied magnetic field of 90~kOe, the magnetization of \pr~and \nd~ is $\rm \approx 1~\mu_B/f.u$ and $\rm \approx 4~\mu_B/f.u$. These values are significantly smaller than the \textit{free-ion} saturation magnetization of g$_J$J = 12.8~$\rm \mu_B/f.u$ (\pr) and 13.1~$\rm \mu_B/f.u$ (\nd), where g$_J$ is the Land\'e g-factor. 
    These reduced values can be attributed to the crystal field effect, as at low temperatures only the low-lying crystal field split levels contribute to the magnetization. Between \pr~and \nd, the magnetization value of $\rm \approx 1~\mu_B/f.u$ for \pr~as against $\rm \approx 4~\mu_B/f.u$ for \nd~ is likely due to the singlet ground state of one of the two crystallographically inequivalent Pr$^{3+}$ ions, as pointed out earlier. This interpretation gains more weight if one notes that in the RP series (where the non-magnetic singlet nature of one of the two Pr$^{3+}$ ions in the structure has been vividly shown), the magnitude of M(H) between Nd and Pr samples have a similar relation~\cite{Dibyata2020}. 
    
    In the inset of Fig.~\ref{MT}(c), the M(H) for \la~ is shown, which is overshadowed in the main panel due to the R$^{3+}$ sublattice. At 5~K, M(H) of \la~ exhibits a ferromagnet-like steep initial increase, but at high fields M(H) continues to increase linearly, showing no signs of saturation. This behavior is qualitatively similar to that previously reported~\cite{Zhang2017}. However, the saturation moment of the ferromagnetic component (obtained by extrapolating the linear part backwards to H = 0) is 0.2~emu~g$^{-1}$ in our sample, but it is close to 0.75~emu~g$^{-1}$ in Ref.~\cite{Zhang2017}. At the same time, dM/dH from the linear part is $\rm 2 \times 10^{-3}~emu~mol^{-1}~Oe^{-1}$ in both the studies. Zhang~\etal~\cite{Zhang2017} argued that these two components actually represent contributions from two different phases. While the saturation is due to the Ni-metal phase, which forms in trace amounts during the reduction process, the linear increase at high fields is intrinsic to the T' phase, which is well supported by the observation that dM/dH has the same value for the two samples. The smaller saturation value for our sample indicates a smaller Ni-metal component in our sample.
    
    Similarly, the M(H) of \pr~samples in some previous studies is reported to exhibit a large ferromagnetic component~\cite{PNASZhang, Huangfu2020}. In the Supporting Information provided with Ref.~\cite{PNASZhang}, the authors delved deeper into the intrinsic versus extrinsic origin of the ferromagnetic component and concluded that their data are difficult to reconcile with the intrinsic argument put forward in Ref.~\cite{Huangfu2020}. They observed that the ferromagnetic component is isotropic, unlike the quantity dM/dH, which exhibits a significant anisotropy between the in-plane and out-of-plane data, which led them to conclude with reasonable certainty that the ferromagnetic component has an extrinsic origin related to the presence of Ni metal, as discussed above. Zhang~\etal~ also showed that the saturation magnetization of the ferromagnetic component is almost temperature independent~\cite{Zhang2017}, which supports the presence of a ferromagnetic impurity in the form of Ni metal. This scenario is highly plausible since reduction beyond the optimized duration results in very rapid decomposition of the compound into La$_2$O$_3$ and Ni metal.

    \subsubsection{Specific heat}
\begin{figure}[t]
		\centering
		\includegraphics[width=\columnwidth]{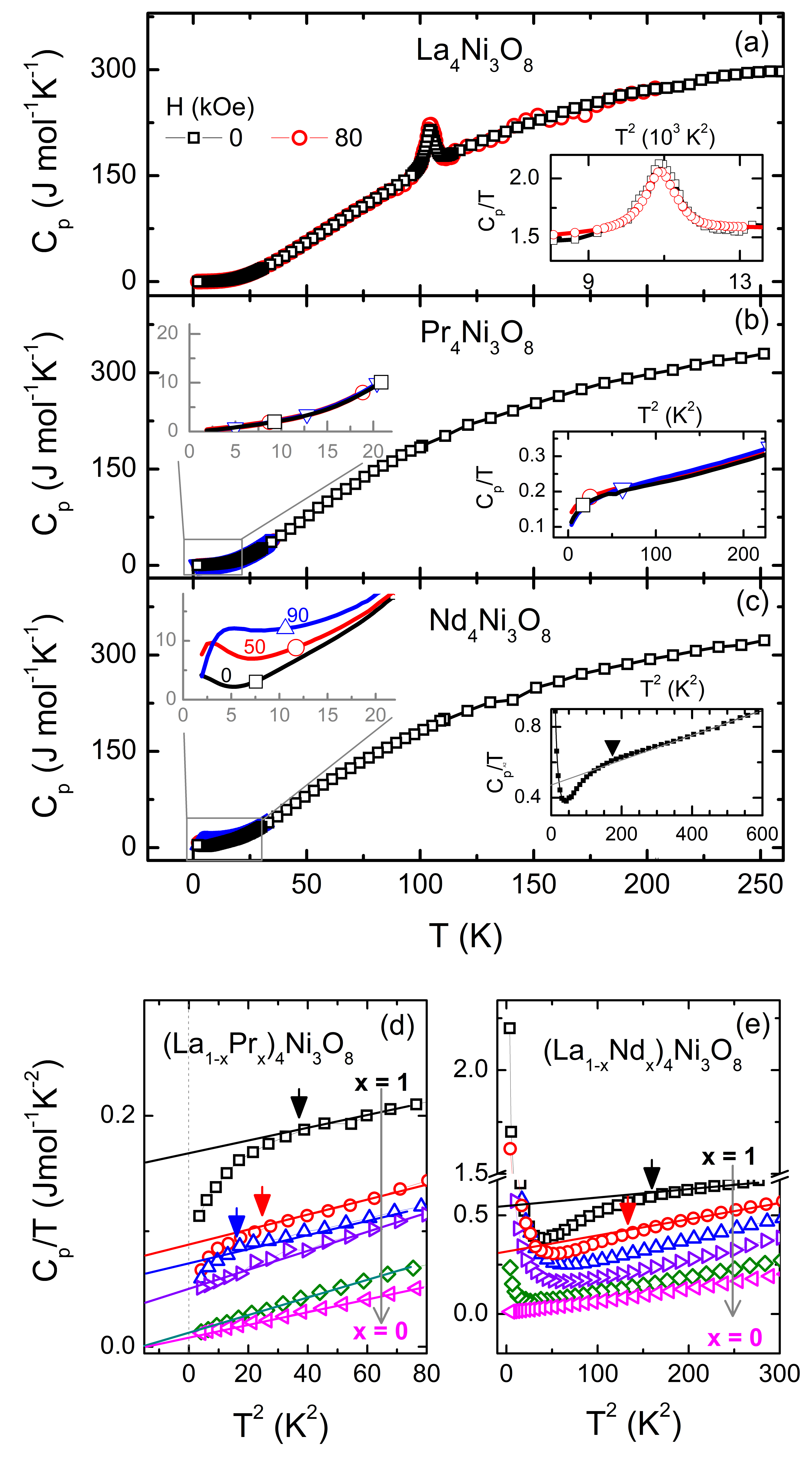}
		\caption {(a), (b), and (c) show the temperature variation of $\rm C_p$ for the \la, \pr~and \nd~samples respectively. The lower insets in panels (a), (b), and (c) show the $\rm C_p/T$ vs T$^2$ plot. In panels (b) and (c) the upper inset shows a zoomed-in view of the low-temperature specific heat. Note that the data points are dropped for clarity. The lines though the data points are a guide to the eye. Panel (d) and (e) show the zero field $\rm C_p/T$ vs $\rm T^2$ plot for the Pr and Nd series, respectively. The arrows pointing downward mark the position of T$^\ast$ for various compositions using the same color as used for the corresponding plot. The legends used in (c) and (d) are the same.}
		\label{CPP}
    \end{figure}

    The molar specific heat (C$_p$) of the parent R$_4$Ni$_3$O$_8$ (R = La, Pr, and Nd) compounds are shown in Fig.~\ref{CPP} (a-c). In~\la, the concomitant charge/spin (CS) stripe ordering is manifested as a peak 
    near \T~$=$~105 K, as shown in Fig.~\ref{CPP}(a). The temperature \T~of the anomaly is in excellent agreement with the previous reports~\cite{PNASZhang, PhysRevLett.104.206403, PRLPressureLa, PRL_NPD_La}, and with the position of the anomaly in the lattice parameters in Fig.~\ref{XRD1}(a1), Fig.~\ref{XRD1}(a2), and Fig.~\ref{XRD1}(a3); or with the position of the anomaly in the temperature variation of M/H for $x$ = 0, as shown in Fig.~\ref{MT}(a). 
    Let us now turn to the specific heat of \pr~and \nd~samples, shown in Fig.~\ref{CPP}(b) and Fig.~\ref{CPP}(c), respectively. The first thing to note is that for both these samples, the temperature variation of C$_p$ is smooth over the entire temperature range. In particular, the 105 K anomaly present for the~\la~sample is absent for these samples, in line with the temperature-dependent PXRD and magnetization data, discussed earlier. Near room temperature, their specific heats are nearly equal, but about 20-25~J~mol$^{-1}$~$\rm K^{-1}$ higher than that of~\la. The excess specific heat is associated with the CF levels of the magnetic rare-earth ions. While the high-temperature specific heat is featureless and varies smoothly, the low-temperature specific heat of these two samples (Pr and Nd) exhibits interesting features that are worth discussing here. We first note that the specific heat of \pr~near 20~K is approximately one-half that of \nd~at the same temperature. Furthermore, upon cooling below about 6~K or so the specific heat of \pr~shows an anomalous behavior, which is more clearly depicted in the lower inset of Fig.~\ref{CPP}(b), where $\rm C_p/T$ is plotted against $\rm T^2$. We note that below a temperature marked as $\rm T^\ast \approx~6~K$, the $\rm C_p/T$ 
    shows a sharp downturn. Interestingly, this feature shows no magnetic field dependence. 

    On the other hand, the low temperatures specific heat of \nd~displays an upturn upon cooling 5~K as shown in the upper inset of Fig.~\ref{CPP}(c). Under an applied magnetic field, the onset temperature for this upturn shifts to higher values resulting in a broad peak at lower temperatures; this peak shifts to higher temperatures as the field strength increases -- a feature typical of the Schottky-type anomaly, which, in this case, could be associated with the Zeeman split lowest crystal field level of Nd$^{3+}$ ions in~\nd. In Nd$_4$Ni$_3$O$_{10}$, a similar behaviour due to the Zeeman splitting of the Kramer's doublet ground state was previously reported~\cite{Dibyata2020}. Interestingly, when plotted as $\rm C_p/T$ versus T$^2$ (see lower inset in Fig.~\ref{CPP}(c)), the specific heat of \nd~also exhibits a sharp downturn, similar to that seen for~\pr, below a temperature of 13~K (this feature is in addition to the upturn due to the Schottky effect). 
    In \pr, the absence of Schottky contribution is consistent with the interpretation that Pr$^{3+}$ has a singlet ground state. The presence of an anomalous downturn below $\rm T^\ast \approx~6~K$ in the Pr and 13~K in the Nd sample is magnetic field independent and does not seem to arise from the rare-earth sublattice. In order to verify this hypothesis, let us examine the specific heat of the samples from the (La, Pr) and (La, Nd) series. 

    The low-temperature specific heat of (La$_{1-x}$R$_x$)$_4$Ni$_3$O$_8$ (R $=$ Pr, Nd) samples is shown in Fig.~\ref{CPP}(d) and Fig.~\ref{CPP}(e), respectively. 
    In the Pr-doped series, with decreasing Pr-doping the magnitude of $\rm C_p/T$ decreases progressively, and the temperature $\rm T^\ast$ also decreases to 5~K for x = 0.9 and 4~K for x = 0.75; however, for x = 0.5, $\rm C_p/T$ varies linearly without any perceptible downturn. Thus, $\rm T^\ast$ for this sample, if non-zero, lies below 2~K (the lowest temperature in our measurements). 
    In the Nd-series, the downturn in $\rm C_P/T$, which was seen below 13~K in the pristine sample, shifts down to 11~K for x = 0.9. However, no such feature could be seen for x = 0.75. If we plot $\rm T^\ast$ versus the average R-site lattice parameter, which is shown later in the manuscript (see Fig.\ref{PD}), one sees a linear behavior as a function of $r_{\Bar{R}}$ across the two series. This suggests that the anomaly $\rm T^\ast$ has a common origin in both \pr~and \nd, and it is controlled by the average R-site ionic radius. Going by this, $\rm T^\ast$ in $x = 0.75$ should fall between 5~K and 6~K, which coincides with the temperature below which the Schottky contribution dominates, masking the weak downturn for this sample.          
    
    We fitted the low temperature C$_p$ for both the (La$_{1-x}$Pr$_x$)$_4$Ni$_3$O$_8$ and (La$_{1-x}$Nd$_x$)$_4$Ni$_3$O$_8$ series using the expression: $\rm \frac{C_p}{T}= \gamma + \beta T^2$, where the coefficients $\gamma$ and $\beta$ represent the electronic and lattice contributions, respectively. For this purpose, we used the data above 6~K in~\pr~series and 13~K~in the~\nd~series. Above these temperatures, the $\rm C{_p}/{T}$ versus $\rm T^2$ plots look fairly linear over the temperature range shown as shown in Fig.~\ref{CPP} (d) and Fig.~\ref{CPP}(e). The estimated $\gamma$ value turns out to be $\rm 113~mJ~mol^{-1}~K^{-2}$ for \pr~and $\rm 485~mJ~mol^{-1}~K^{-2}$ for~\nd. These values are unusually high, and one may be inclined to conclude that these samples possibly exhibit a heavy fermion behavior. However, before arriving at such a conclusion, one has to first carefully discard the crystal field contribution arising from the low-lying crystal field split levels of R$^{3+}$ ions. This is a complex task, requiring the knowledge of the crystal field splitting scheme. 
    In the case of \la, the fitting was performed in the temperature range below 7~K, which resulted in $\rm \gamma \approx 10~mJ~mol^{-1}~K^{-2}$
    which is comparable to that for La$_4$Ni$_3$O$_{10}$~\cite{Dibyata2020} that exhibits a metallic ground state. Since the ground state of \la~is insulating, the fairly high $\gamma$ value for this sample can arise from the spin-wave contribution associated with the long-range ordering of the Ni moments. 

	
    


    \subsubsection{Electrical Transport}
\begin{figure}[t!]
		\centering
		\includegraphics[width= \columnwidth]{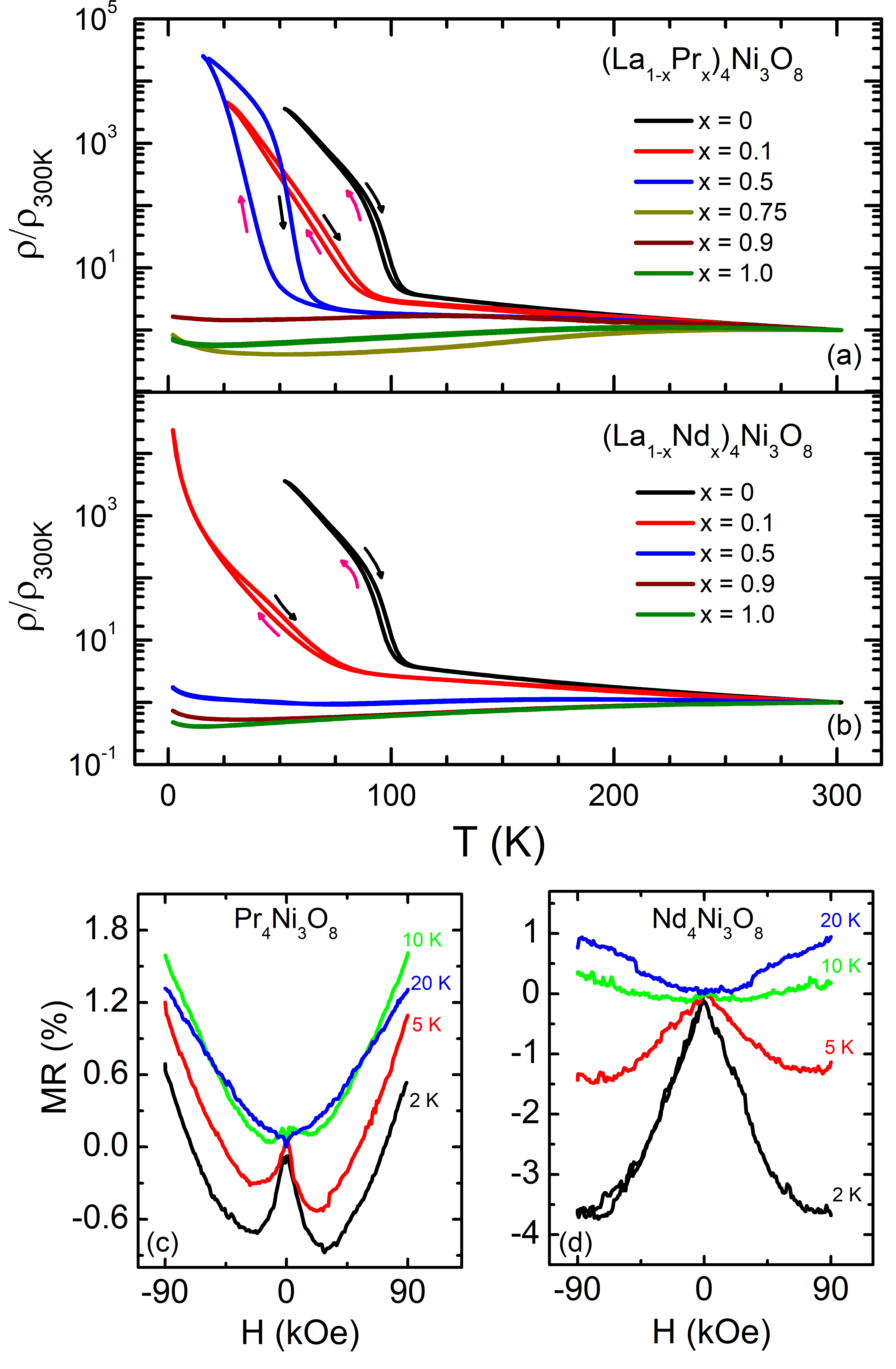}
		\caption {(a) and (b) show the value of $\rm \rho(T)$/$\rho_{300K}$ plotted as a function of temperature for La$_{1-x}$Pr$_x$Ni$_3$O$_8$ and La$_{1-x}$Nd$_x$Ni$_3$O$_8$ (0 $\leq$ x $\leq$ 1.0) samples, respectively. The arrows represent cooling and heating curves. Magnetoresistance of (c) \pr~and, (d) \nd~at T = 2, 5, 10, and 20~K under an applied magnetic field ranging from -90~kOe to 90~kOe.}
		\label{Rho}
    \end{figure}

The temperature-dependent normalized electrical resistivity, $\rm \rho/\rho_{300 K}$ where $\rm \rho_{300 K}$ is the resistivity at 300~K, of (La$_{1-x}$Pr$_x$)$_4$Ni$_3$O$_8$ and the (La$_{1-x}$Nd$_x$)$_4$Ni$_3$O$_8$ samples is shown in Fig.~\ref{Rho}(a) and Fig.~\ref{Rho}(b), respectively. The MIT associated with the CS-stripe ordering is clearly captured in \la~near \T~= 105 K, in good agreement with previous reports~\cite{PNASZhang, PhysRevLett.104.206403, PRLPressureLa, Zhang2017}. Below this temperature, the resistivity shows a sharp increase upon cooling. In the (La$_{1-x}$Pr$_x$)$_4$Ni$_3$O$_8$ samples, the MIT is suppressed with the increase in the Pr-doping. The \T~in the Pr-doped samples agrees fairly well with anomalies in the lattice parameter and magnetization data presented in the previous sections. In the Pr-rich samples, $x = 0.9$ and $x = 0.75$, the sharp increase in resistivity, characteristic of the MIT is not seen, indicating that the critical doping required for the complete suppression lies in the range 0.5 $\rm \leq x \leq 0.75$, in agreement with Ref.~\cite{Xinglong_CM}. Note that in Ref.~\cite{Xinglong_CM}, the chemical formula is written as (Pr$_{1-z}$La$_z$)$_4$Ni$_3$O$_8$, and \T~is shown to disappear between $\rm 0.6 \leq 1-z \leq 0.7$, which is contained within the interval found here. Although no MIT is seen for x = 0.9 and 0.75, the $\rho(T)$ exhibits a shallow minimum, with the slope $\rm d\rho(T)/dT$ becoming steeper at very low temperatures. It should be pointed out that the resistivity of the pristine sample also behaves similarly, ruling out any extrinsic origin or an origin related to sample inhomogeneity. 
This observation is consistent with several previous reports~\cite{miyatake4080321chemical, Zhang2017, NdSmSulphur, Nakata2016, Pan2022}. 
The metallic behavior of \pr~is in agreement with the theoretical calculations~\cite{PhysRevB.102.020504, Pr438nNdNiO2, Review2022}, where the presence of a large hole pocket contribution from the Ni $d_{x^2-y^2}$ band at the Fermi level contributes to the charge transport. 
The residual resistivity ratio (RRR) for our metallic Pr-doped samples (i.e., 0.75 $\leq$ x $\leq$ 1) varies from 1 to 2. Such low values of RRR for these samples could be due to the presence of microcracks and stacking faults (see Fig.~\ref{TEM}(b)) which may have appeared during the reduction process. Similar behavior is also seen for the La-Nd series albeit with a difference that in this case a clear MIT is seen only for $x = 1$ and $0.75$. 
The upturn in the low-temperature resistivity of the Pr or Nd-doped metallic samples can arise due to a variety of reasons, including weak localization~\cite{WL}, electron-electron interaction or Kondo-like spin-dependent scattering mechanism~\cite{Kondo}. To understand, we tried fitting the low-temperature data to a logarithmic temperature dependence (lnT) as well as to a T$^{0.5}$ dependence (see Fig. S6 in Supplementary Material). Both scenarios result in fits that do not look completely satisfactory, hence making it difficult to assign either of the two scenarios for explaining the resistivity upturn. 

    \textit{Magnetoresistance}: The magnetoresistance (MR) refers to the change in resistance of a material under the influence of an applied magnetic field and is defined as, MR (\%) = $[(\rho(H) - \rho(0))/\rho(0)]\times 100$, where H is the externally applied magnetic field; $\rho(H)$ and $\rho(0)$ are the resistivities (or resistance) of the sample in the presence and absence of magnetic field, respectively. The MR data for the \pr~and \nd~samples is shown in Fig.~\ref{Rho}(c) and Fig.~\ref{Rho}(d), respectively. At 10 K and 20 K, the MR of \pr~is positive over the whole temperature range. The field dependence is parabolic, except close to the origin, where the variation is linear. 
    The quadratic dependence arises due to the extra scattering that the carriers encounter as they move under the influence of the Lorentz force acting on them. 
    At lower temperatures of 5~K and 2~K, the MR first decreases linearly up to 20-30~kOe and thereafter increases quadratically, becoming positive at higher fields. The low-field linear region is a typical signature of the phenomenon known as weak antilocalization (WAL)~\cite{HLN, Xing2013, WALgraphene, WAL_Na2IrO3}. The effect is overcome at high fields, restoring the quadratic behavior. 
    
    For the \nd~sample a negative MR of nearly 4\% has been observed at 2~K. The MR initially increases with the field before saturating at higher fields to a value close to 4\% under a maximum applied field of $\pm$90~kOe. The magnitude of negative MR decreases as the temperature is increased from 2~K to 5~K. Thereafter, the MR shows a crossover from negative to small positive MR at 10~K and 20~K. We believe that the negative MR in \nd~sample at low temperatures is related to spin-disorder scattering~\cite{SpinDisorder}. Recall the presence of Schottky anomaly in the low-temperature specific heat of \nd. This Schottky anomaly is due to the Zeeman splitting of the lowest crystal field split level. At low temperatures, as the magnetic field strength increases, more and more Nd-ions occupy the lower Zeeman energy level; or, in other words, with increasing field strength more and more Nd moments will line up along the field direction. This reduces the spin-disorder scattering, leading to a negative MR, as seen here. The MR saturates at high fields when the Nd-moments are all more or less aligned along the field direction. The MH curve at 2~K in Fig.~\ref{MT}(d), shows that the magnetization nearly saturates at 2~K under 90 kOe. An alternative possibility is that the negative MR in \nd~also arises due to the weak anti-localization (WAL) effect as discussed above in the context of \pr. However, the saturation of MR at high magnetic fields rules out this scenario. 

\subsection{Phase diagram}
\begin{figure}[b]
    \centering
    \includegraphics[width=1\linewidth]{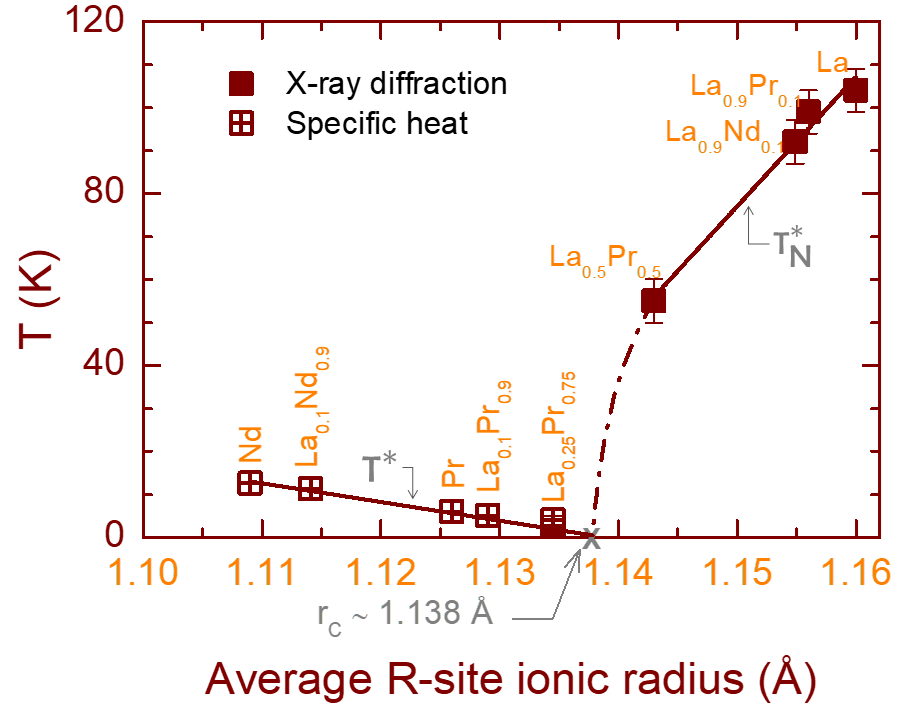}
    \caption{The variation of \T~and T$^\ast$ as a function of average R-site ionic radius ($r_{\Bar{R}}$). $\rm r_c$ denotes the critical value of $\rm r_{\Bar{R}}$ at which \T~($\rm r_{\Bar{R}}$~$>$~$\rm r_c$) and T$^\ast$ ($\rm r_{\Bar{R}}$~$<$~$\rm r_c$) tends to zero.}
    \label{PD}
\end{figure}

    Having analyzed the temperature dependence of powder X-ray diffraction data and various other physical quantities, we have enough information to draw a phase diagram depicting the variation of \T~ (the concomitant charge-spin stripe ordering and metal-to-insulator transition) and T$^\ast$ (the low-temperature specific heat anomaly) and the phases bounded by them. Table.~\ref{Table_PD} summarizes the values of \T~ and T$^\ast$ for various values of $\rm r_{\Bar{R}}$ calculated as $\rm r_{\Bar{R}} = (1-x)r_{La} + xr_{Pr/Nd}$, where $\rm r_{R}$ is the ionic radius of R$^{3+}$ for R = La, Pr, and Nd, in the eight-fold coordination and $\rm \Bar{R} \equiv La_{1-x}Pr_x~or~La_{1-x}Nd_x$. The phase diagram is shown in Fig.~\ref{PD}. With decreasing $\rm r_{\Bar{R}} $, the transition temperature \T~decreases almost linearly down to $\rm r_{\Bar{R}} = 1.143~\AA$, i.e., the composition $\rm \Bar{R}$~$\equiv$~La$_{0.5}$Pr$_{0.5}$. However, in the composition $\rm \Bar{R} \equiv La_{0.5}Nd_{0.5}$, which corresponds to $\rm r_{\Bar{R}}  = 1.1345~\AA$, no MIT could be detected either in x-ray diffraction or other physical properties. This suggests that \T~vanishes in the range $\rm 1.134~\AA < r_{\Bar{R}} < 1.143~\AA$, this agrees well with Ref.~\cite{Xinglong_CM}, where this range is even better defined as $\rm 1.136~\AA < r_{\Bar{R}} < 1.140~\AA$. Thus, \T~line drops quite suddenly to 0 once the Pr-doping in \la~exceeds a critical value in this range. The temperature T$^\ast$ has been identified in this study and has no reference in the previous literature to the best of our knowledge. Since, T$^\ast$ (the temperature below which $\rm C_P/T$ shows a sharp downturn) is present for both \pr~and \nd, and their intermediate compositions, and as this transition also scales almost linearly with $\rm r_{\Bar{R}} $, we have good reasons to argue that this characteristic temperature is associated with the Ni $d$ electrons. The variation of T$^\ast$ with $\rm r_{\Bar{R}} $ is shown in the phase diagram in Fig.~\ref{PD}. In the limit T$^\ast \rightarrow 0$, the value of $\rm r_{\Bar{R}}$, which we have labeled as $\rm r_c$, comes out to be 1.138~\AA, which lies well within the range $\rm 1.136~\AA < r_{\Bar{R}} < 1.140~\AA$ where \T~vanishes. Hence, these two temperature scales seem to be mutually exclusive; or, in other words, coming from the Nd or Pr side of the phase diagram, one can argue that the electronic instability associated with the phenomenon underlying $\rm T^\ast$ prevents the MIT from setting in.    

\begin{table}[t]
    \setlength{\tabcolsep}{16pt}
    \caption{The transition temperatures \T~and T$^\ast$ in degree Kelvin (K) for various samples. The quantity $\rm r_{\Bar{R}} $ is the average R-site ionic radius measured in \AA.}
		\label{Table_PD}
		\begin{center}
		\begin{tabular}{ c    c    c    c  }
		\hline
            \hline
            \\
            $\Bar{R}$ & $\rm r_{\Bar{R}}$ & \text{T} & $T^\ast$
            \\
		\hline
            \hline
		La & 1.16  & 104 & - \\
		\hline
		La$_{0.9}$Pr$_{0.1}$ & 1.156 & 99 & - \\
		\hline
		La$_{0.9}$Nd$_{0.1}$ & 1.155 & 92 & - \\
		\hline
		La$_{0.5}$Pr$_{0.5}$ & 1.143 & 55 & - \\
		\hline
		La$_{0.5}$Nd$_{0.5}$ & 1.134 & 0 & -  \\
		\hline
            La$_{0.25}$Pr$_{0.75}$ & 1.134 & - & 4 \\
		\hline
		La$_{0.1}$Pr$_{0.9}$ & 1.129 & - & 5  \\
		\hline
            Pr & 1.126 & - & 6  \\
		\hline
            La$_{0.25}$Nd$_{0.75}$ & 1.122 & - & -  \\
		\hline
            La$_{0.1}$Nd$_{0.9}$ & 1.114 & - & 11.2 \\
		\hline
            Nd & 1.109 & - & 12.6  \\
		\hline
	    \end{tabular}
		\end{center}
	\end{table}

    \section{Summary \& Conclusions}
    \label{SC}

\la~is an interesting system, showing concomitant charge-spin stripes ordering and metal-to-insulator transition near \T~=105~K. Intriguingly, this transition does not occur in its \pr~and \nd~analogues, both of whom show a metallic behavior. Here, we examined the effect of progressively decreasing the R-site ionic radius by doping at the La site with Pr and Nd separately, leading to two sets of samples: (La$_{1-x}$Pr$_x$)$_4$Ni$_3$O$_8$ and (La$_{1-x}$Nd$_x$)$_4$Ni$_3$O$_8$, $\rm 0 \leq x \leq 1$. By combining the samples from the two doping series, we show that the transition temperature \T~initially decreases linearly as the average R-site ionic radius ($r_{\Bar{R}}$) decreases, and then vanishes suddenly in the narrow range $\rm 1.134~\AA \leq r_{\Bar{R}} \leq 1.143~\AA$, in agreement with Ref.~\cite{Xinglong_CM} where this range is even better defined as $\rm 1.136~\AA \leq r_{\Bar{R}} \leq 1.140~\AA$. A careful examination of the low-temperature specific heat revealed the presence of a new energy scale characterized by the temperature $\rm T^\ast$ below which $\rm C_P/T$ exhibits a sharp drop. The variation of $\rm T^\ast$ with $\rm r_{\Bar{R}}$ for various intermediate compositions, including~\pr~($\rm T^\ast$ $\sim$ 6~K)~and~\nd~($\rm T^\ast \sim 13 K$), we found that: (a) $\rm T^\ast$ versus $r_{\Bar{R}}$ plot follows a linearly decreasing trend with increasing $\rm r_{\Bar{R}}$, vanishing at $\rm r_{\Bar{R}}~\approx~1.138~\AA$, which lies exactly in the range where \T~disappears with decreasing $\rm r_{\Bar{R}}$. In other words, we demonstrate that the sudden disappearance of \T~ upon decreasing $\rm r_{\Bar{R}}$ is associated with the emergence of a new phase below $\rm T^\ast$, where $\rm T^\ast$ increases with decreasing $\rm r_{\Bar{R}}$ in the region $\rm r_{\Bar{R}}$~$<$~$\rm r_c$. Up to now, more advanced tools such as ARPES and Neutron have been used primarily to understand the nature of phase transition at~\T. We propose that for a deeper understanding of the nature of \T~and its peculiar dependence on the R-site ionic radius (its sharp disappearance below a certain critical $\rm r_{\Bar{R}}$~$=$~$\rm r_c$), low-temperature experiments unraveling the nature of the new phase below $\rm T^\ast$, as \T~vanishes will be very useful. As $\rm T^\ast$ is controlled by $\rm r_{\Bar{R}}$, seen for both~\pr~and \nd~(i.e., independent of the choice of R), one can argue that this transition does not originate from the rare-earth sublattice. Given the fact that Pr magnetism is weakened by the crystal field splitting due to non-magnetic singlet state, the prospects of studying the nature of the new phase below $\rm T^\ast$ are highly favorable in \pr~and Pr-rich samples.
\\
\section*{Acknowledgments}
	 SS acknowledges financial support from SERB (WMR/2016/003792). DR acknowledges financial support from IEEE Magnetics Society Educational Seed Funding. We are thankful to beamline scientists Francois Fauth, Catalin Popescu, and Aleksandr Missiul at the MSPD-BL04 beamline at ALBA Synchrotron facility. SS and DR would like to acknowledge the Department of Science and Technology, India (Grant No. SR/NM/Z-07/2015) for providing financial support for carrying out the Synchrotron	experiments at ALBA, and to the Jawaharlal Nehru Centre for Advanced Scientific Research (JNCASR) for facilitating it. SS and DR would like to thank UGC DAE Indore for providing PPMS facility to carry out magnetotransport and magnetization experiments.

	\bibliography{RefNi}
 
 \renewcommand{\thefigure}{S\arabic{figure}}
 \newpage
 \onecolumngrid
 \appendix
 \setcounter{figure}{0}
 \onecolumngrid

 \section*{Supplementary Information: Investigating the cause of crossover from charge/spin stripe insulator to correlated metallic phase in layered T' nickelates -R$_4$Ni$_3$O$_8$}
\begin{figure}[htbp]
		\centering
		\includegraphics[width= 0.9\textwidth]{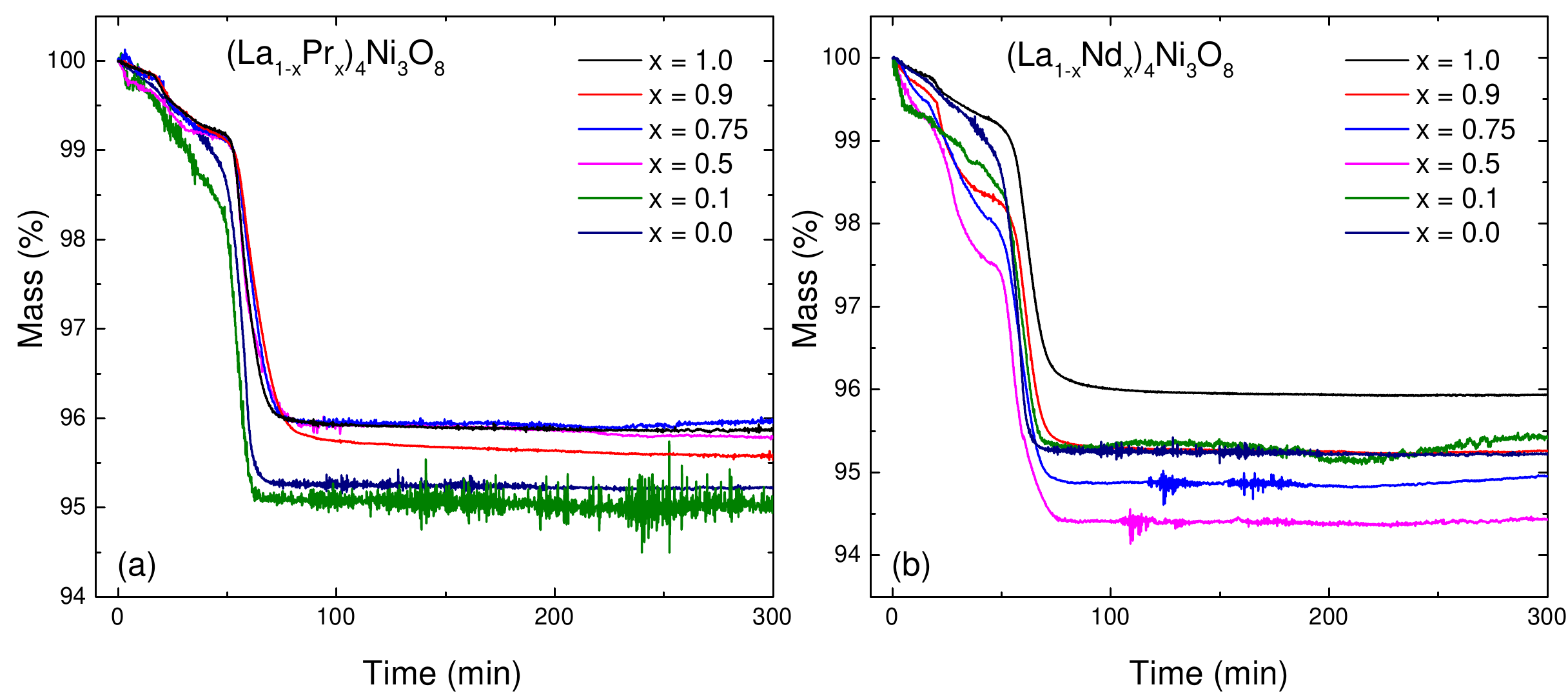}
		\caption {(a) and (b) show the TGA isotherms carried out at 600$^\circ$C in Ar-H$_2$ (10\%) atmosphere for (La$_{1-x}$Pr$_x$)$_4$Ni$_3$O$_8$ and (La$_{1-x}$Nd$_x$)$_4$Ni$_3$O$_8$ samples respectively. }
		\label{S1}
	\end{figure}

    \maketitle	
    \paragraph{Thermogravimetric analysis}~A complete decomposition of the samples was carried out by conducting isotherm runs at 600$^\circ$C using 10\% Ar - H$_2$ atmosphere at a heating rate of 10 K/min in a high resolution TGA setup (Netzsch STA 449 F1) as shown in Fig.\ref{S1}(a) and (b). From these experiments, we determined the oxygen stoichiometry of all the samples as shown in Table~\ref{OC}.\\

    	\begin{table*}[htbp]
		\setlength{\tabcolsep}{10pt}
		\caption{Oxygen content of the doped samples estimated using TGA setup.}
		\label{OC}
		\begin{center}
			\begin{tabular}{ |c | c | c | c |c |c |c | }
				\hline
				x in (La$_{1-x}$R$_x$)$_4$Ni$_3$O$_{8\pm\delta}$, R $=$ (Pr and Nd) & Method of synthesis & $\delta$ (LP series) & $\delta$ (LN series) \\
				\hline
				0 & using CaH$_2$ & + 0.29 & + 0.29 \\
				\hline
				0.1 & using CaH$_2$ & + 0.40 & + 0.27  \\
				\hline
				0.5 & Ar-H$_2$ (10\%) & + 0.04 & + 0.72  \\
				\hline
				0.75 & Ar-H$_2$ (10\%) & - 0.05 & + 0.51 \\
				\hline
				0.9 & Ar-H$_2$ (10\%) & + 0.14 & + 0.32  \\
				\hline
				1.0 & Ar-H$_2$ (10\%) & - 0.004 & - 0.13  \\
				\hline
			\end{tabular}
		\end{center}
	\end{table*}

	\begin{figure}[htbp]
		\centering
		\includegraphics[width= 0.9\textwidth]{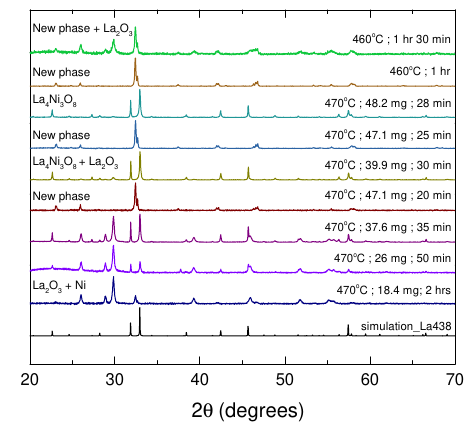}
		\caption {The various reduction trials carried out on the parent La$_4$Ni$_3$O$_{10}$ sample in U.H.P. H$_2$ to obtain the La$_4$Ni$_3$O$_8$ phase. The isotherm time and temperature and sample weight are mentioned alongwith the xrd pattern obtained after the reduction process. The black curve denotes the simulated pattern for La$_4$Ni$_3$O$_8$ in the tetragonal space group of $I4$$\slash$$mmm$.}
		\label{S2}
	\end{figure}

     \begin{figure}[htbp]
		\centering
		\includegraphics[width=\textwidth]{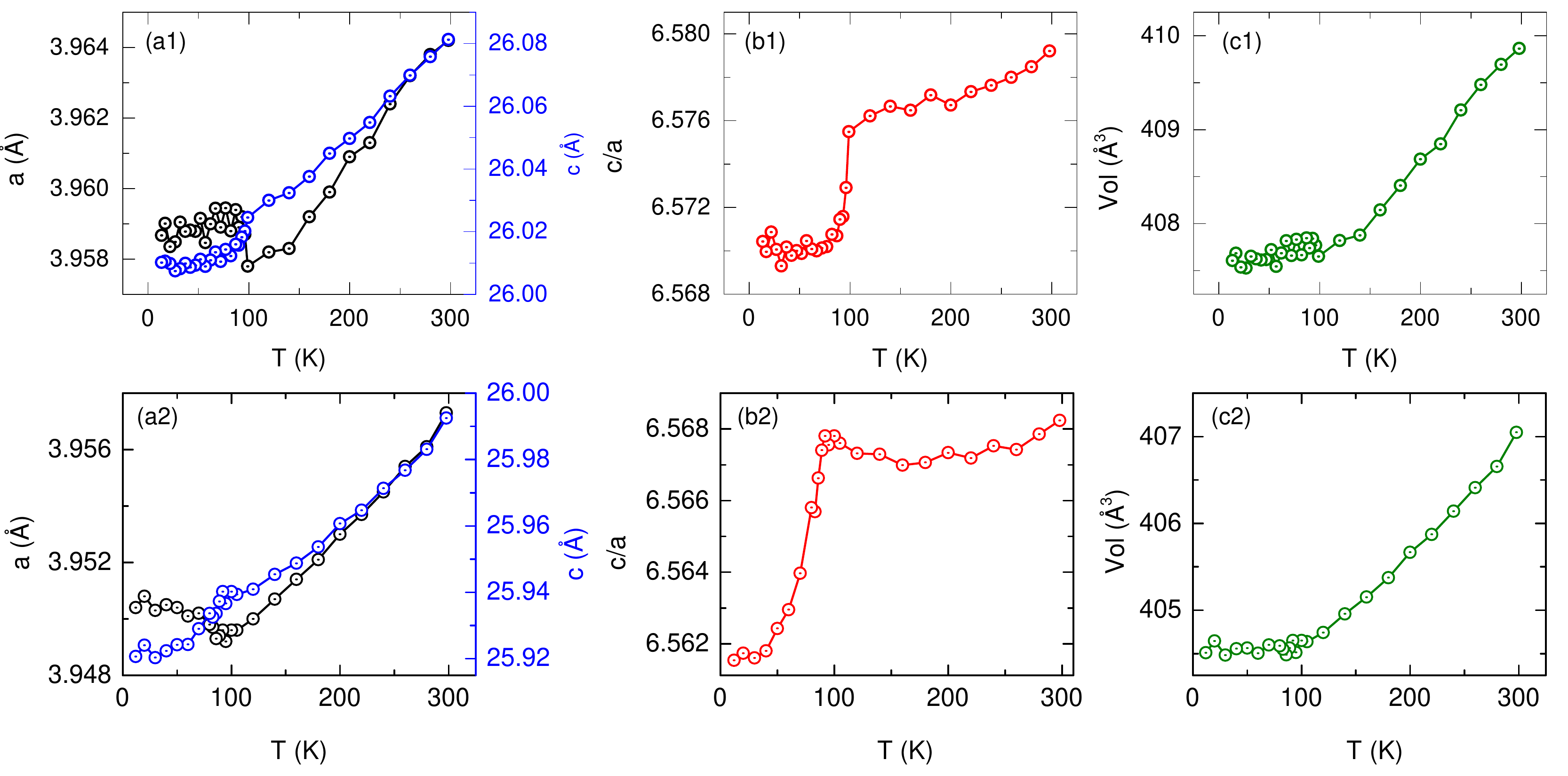}
		\caption{Panels (a1 - c1) and (a2 - c2) show the lab based low temperature XRD data for La$_{3.6}$Pr$_{0.4}$Ni$_3$O$_8$ and La$_{3.6}$Nd$_{0.4}$Ni$_3$O$_8$ samples respectively.}
		\label{XRDLT}
	\end{figure}

	\begin{figure}
		\centering
		\includegraphics[width= \textwidth]{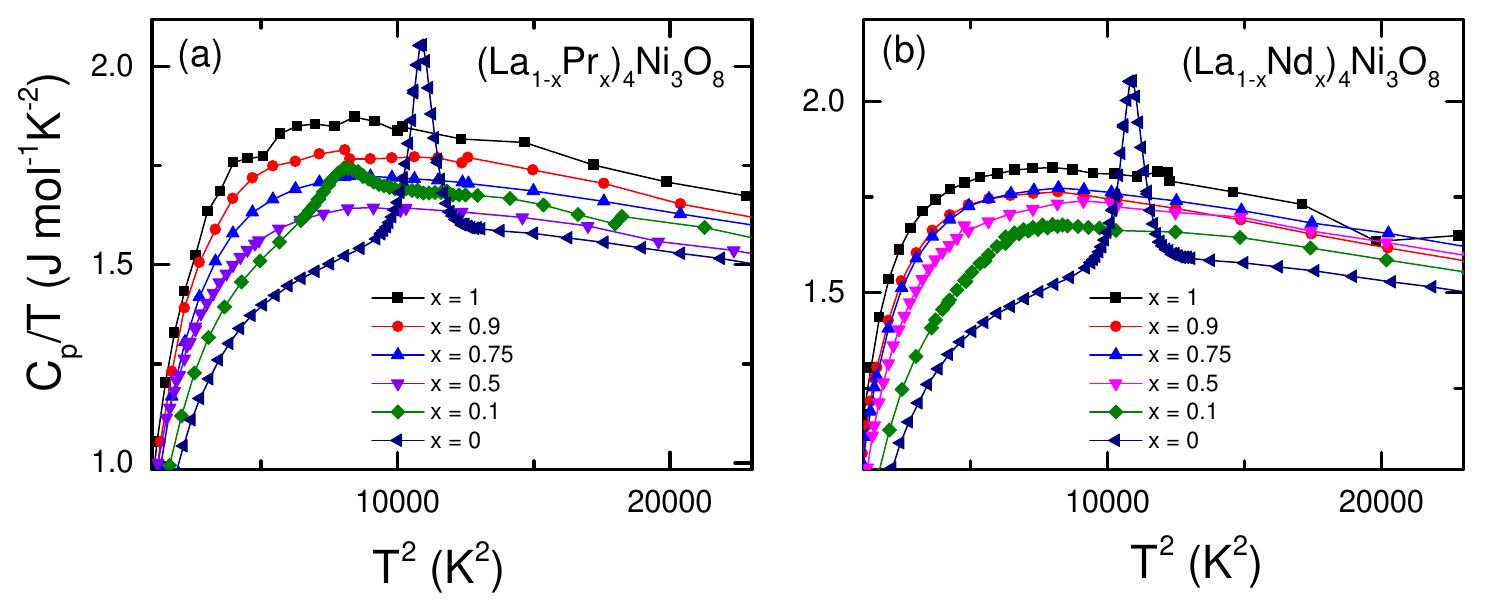}
		\caption {(a) and (b) show the C$_P/T $ vs T$^2$ plot for the (La$_{1-x}$Pr$_x$)$_4$Ni$_3$O$_8$ and (La$_{1-x}$Nd$_x$)$_4$Ni$_3$O$_8$  series respectively.}
		\label{S5}
	\end{figure}

    \begin{figure}
		\centering
		\includegraphics[width= \textwidth]{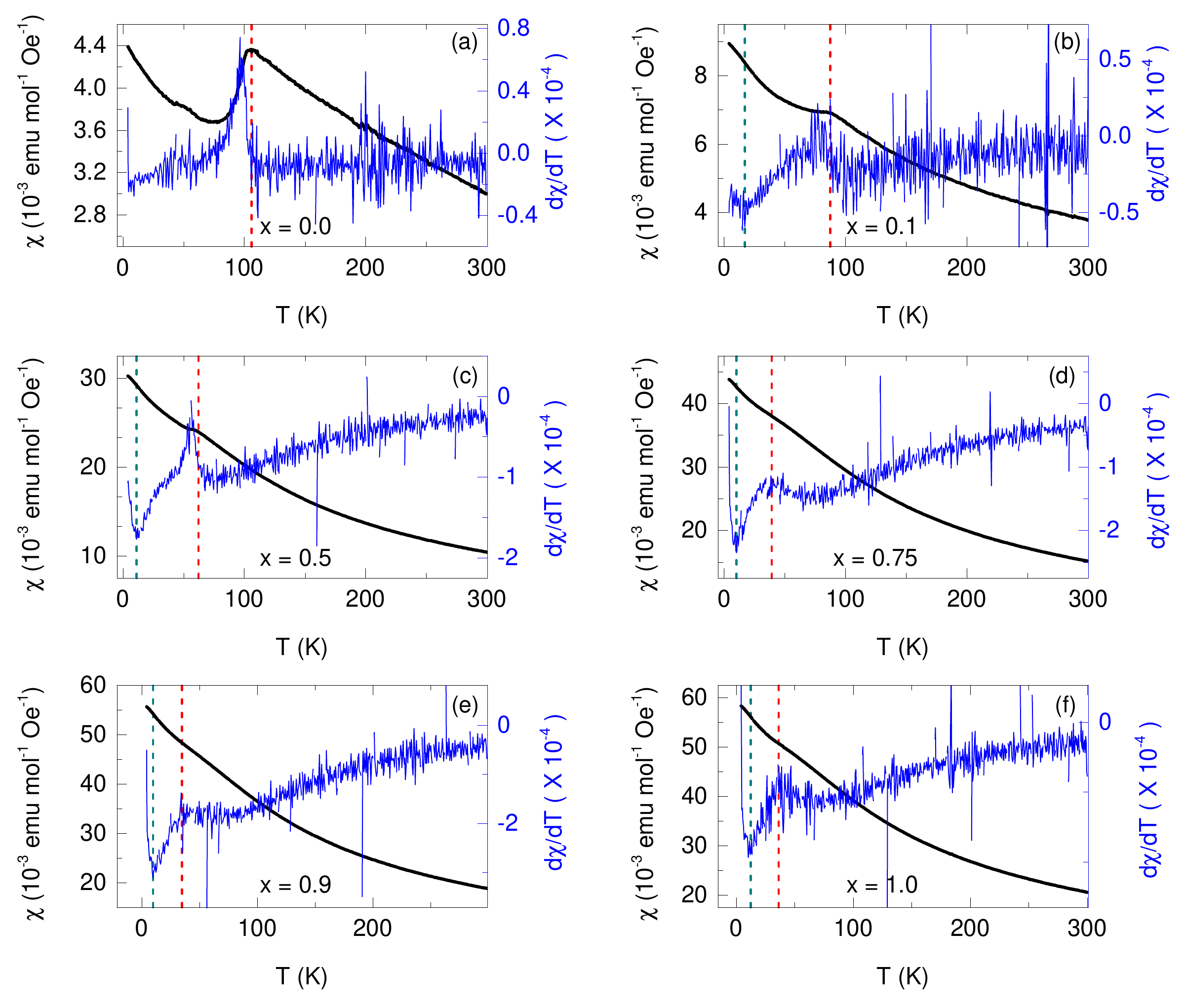}
		\caption {Magnetic susceptibility of the (La$_{1-x}$Pr$_x$)$_4$Ni$_3$O$_8$ series for 0 $\leq$ x $\leq$ 1.0 alongwith its first order derivative (shown in blue). The red dotted line marks the position of the sharp peak in d$\chi$/dt while the dark cyan dotted line marks the low temperature anomaly.}
		\label{S6}
	\end{figure}

    \begin{figure}
		\centering
		\includegraphics[width= \textwidth]{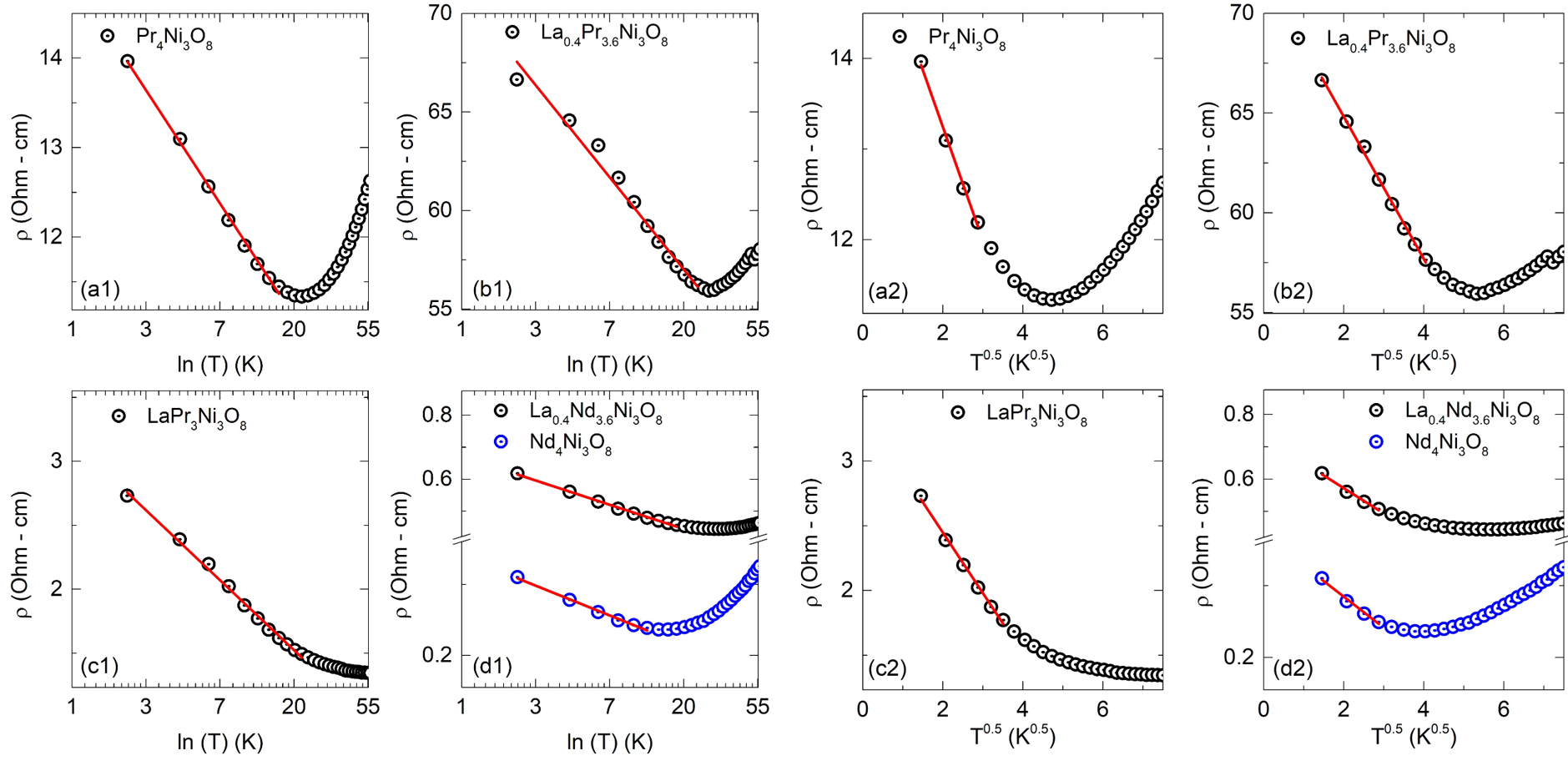}
		\caption {(a1-d1) show the temperature variation of resistivity on a ln(T) scale while (a2-d2) show the temperature variation of resistivity on a T$^{0.5}$ scale for the same set of samples.}
		\label{S4}
	\end{figure}

     \twocolumngrid
 
    \enddocument